\documentclass[a4paper,11pt]{article}
\usepackage[position=t,singlelinecheck=off]{subfig}
\usepackage{amssymb,amsmath}
\usepackage{graphics,graphicx,float}
 \usepackage{mathabx}
\usepackage{epstopdf}
\usepackage[affil-it]{authblk}
\def\beq{\begin{equation}}
\def\eeq{\end{equation}}
\def\bea{\begin{eqnarray}}
\def\eea{\end{eqnarray}}
\def\nn{\nonumber}
\def\lo{\left(}
\def\rc{\right)}
\def\ls{\left[}
\def\rs{\right]}
\def\lg{\left\lgroup}
\def\rg{\right\rgroup}
\makeatletter
 
  \def\@cite#1#2{${\mbox{#1\if@tempswa , #2\fi}}$}
\makeatother

\renewcommand{\theequation}{\arabic{section}.\arabic{equation}}

\title{\LARGE{\sf{Nonclassical  states of light induced via  measurement in a bimodal system}}}
\author{R. Chakrabarti$^{1\, \dag}$, S.V. Iyengar$^{2\, \ddag}$, B.V. Jenisha$^{3\,\S}$}
\affil{${}^{1}$Department of Theoretical Physics, University of Madras, Chennai, 600 025, India,\\
${}^{2}$Department of Data Sciences and Analytics, School of Social Sciences, M S Ramaiah University of Applied Sciences,
Bangalore, 560 054, India,\\
 ${}^{3}$Department of Physics, Government College of Engineering, Srirangam, Tamil Nadu, 620 012, India,\\
$\dag$ ranabir@imsc.res.in \footnote{Corresponding Author}  ,  $\ddag$ sudhaeinstein@gmail.com, $\S$ bvjenisha@gces.edu.in}
\date{}
\begin{document}
\maketitle
\begin{abstract}
We investigate generation of nonclassical photon states via conditional measurement process in a two mode coupled waveguide. Interaction of the fields takes place in a waveguide
beamsplitter due to the overlap between normal modes supported therein.  A quadratic Hamiltonian of  two degrees of freedom describes the hopping interaction. An initial  two mode squeezed state undergoes a unitary evolution governed by the interaction Hamiltonian for a specified time. Following this the bipartite state is subjected to a projective measurement that detects  $n$-th Fock state in one subsystem. The post-measurement  excitation  rendered in the residual subsystem depends on the prior time of interaction between the modes as well as the interaction strength. The Wigner quasiprobability distribution of an arbitrary post-selection state is computed. Its  nonclassicality is examined via the negativity of  the Wigner distribution. The sub-Poissonian nature of the photon statistics is revealed by the Mandel parameter. The  dynamically generated squeezing is evidenced in the post-measurement state. In the ultrastrong coupling regime the parity even and odd states display markedly \textit{different} nonclassical properties. The nonclassicality of the 
post-measurement states obtained here may be \textit{controlled} by varying the interaction strength and the time span  of interaction between the modes.
\end{abstract}

\section{Introduction}
\label{introduce}
Conditional measurements project entangled states of an interacting, say, bipartite system onto a specified state of one subsystem, and thereby trigger emergence of a possible nonclassical state in the complementary subsystem [\cite{Garraway1994}-\cite{WAB2019}]. For a composite system comprising of a quantized  field residing in a cavity and interacting with a two level atom therein, it has been  observed [\cite{Garraway1994}-\cite{WAB2019}] that the field states are notably dependent on the outcome of measurement on the atom after it exits the cavity.
 In particular, markedly nonclassical  field states characterized by sub-Poissonian statistics [\cite{GG1997}, \cite{WAB2019}], and negative  values of Wigner phase space distribution [\cite{A2013}]  have been noticed [\cite{WAB2019}] following selective measurement performed on the atom after it departs the interaction  cavity. Domains of the parametric space that facilitate the post-measurement cavity mode displaying  prominent nonclassical characteristics have been identified  [\cite{WAB2019}]. Moving further, it has been pointed out [\cite{GG1997}] that successive measurements realized on  atoms sequentially passing through the cavity field can amplify the sub-Poissonian properties and a number state of the field mode can be approached as a limit.

\par

In another context  tunneling of photons between two coupled Bosonic modes have been recently examined  [\cite{LA2013}].  This can be demonstrated, for instance, via quantum photonic circuits realized by using integrated silica waveguides on a silicon chip  [\cite{PCRO2008}, \cite{Perets2008}], coupled resonators based on nonlinear optics that have enabled generation of single, bi- and multi-entangled photons  [\cite{Caspani2017}], and optomechanical devices that function parallel to a two level laser system [\cite{GLPV2010}]. Specifically it has been observed  [\cite{LA2013}] that in the strong detuning regime where the tunneling between the two modes is weak, a  significant enhancement in quantum tunneling materializes in the  presence of  prior photons in the terminal mode, whereas 
 photon inhibited tunneling may be produced in the presence of a dominant hopping energy. The effect owes its origin  to the Bosonic nature of the quanta. This suggests that interesting nonclassical many-particle Bosonic states can be generated via conditional measurements in, say,  coupled waveguide devices. In a related feature, it is well-known that the nonclassicality originating from superposition of quantum macroscopic states are of crucial interest [\cite{FSG2015}, \cite{LNA2015}]. Concretely, the two mode squeezed vacuum state has been advanced [\cite{Oudot2015}] towards analyzing quantum effects at macroscopic dimensions.   Experimental demonstration of two mode squeezed states have been recently furnished
[\cite{Leong2023}] by utilizing atoms in a two dimensional optical lattice as quantum registers. Moreover,  using an effective Hamiltonian approach a bimodal cavity structure has been employed [\cite{CBFT2018}] for mapping a quantum state of one cavity mode onto the other mode on demand. Generation of large amplitude cat states [\cite{A2013}] by performing a photon number measurement in one mode of a two mode Gaussian state has been considered [\cite{TYAEF2021}]. Examining a measurement based scheme in a system comprising of atoms interacting with waveguide modes conditional generation of non-Gaussian states such as kitten [\cite{A2013}] states has been obtained [\cite{PB2024}].

\par

\begin{figure}
\begin{center}
\captionsetup[subfigure]{labelfont={sf}}
\includegraphics[scale=0.5,trim= 50 20 30 38,clip]{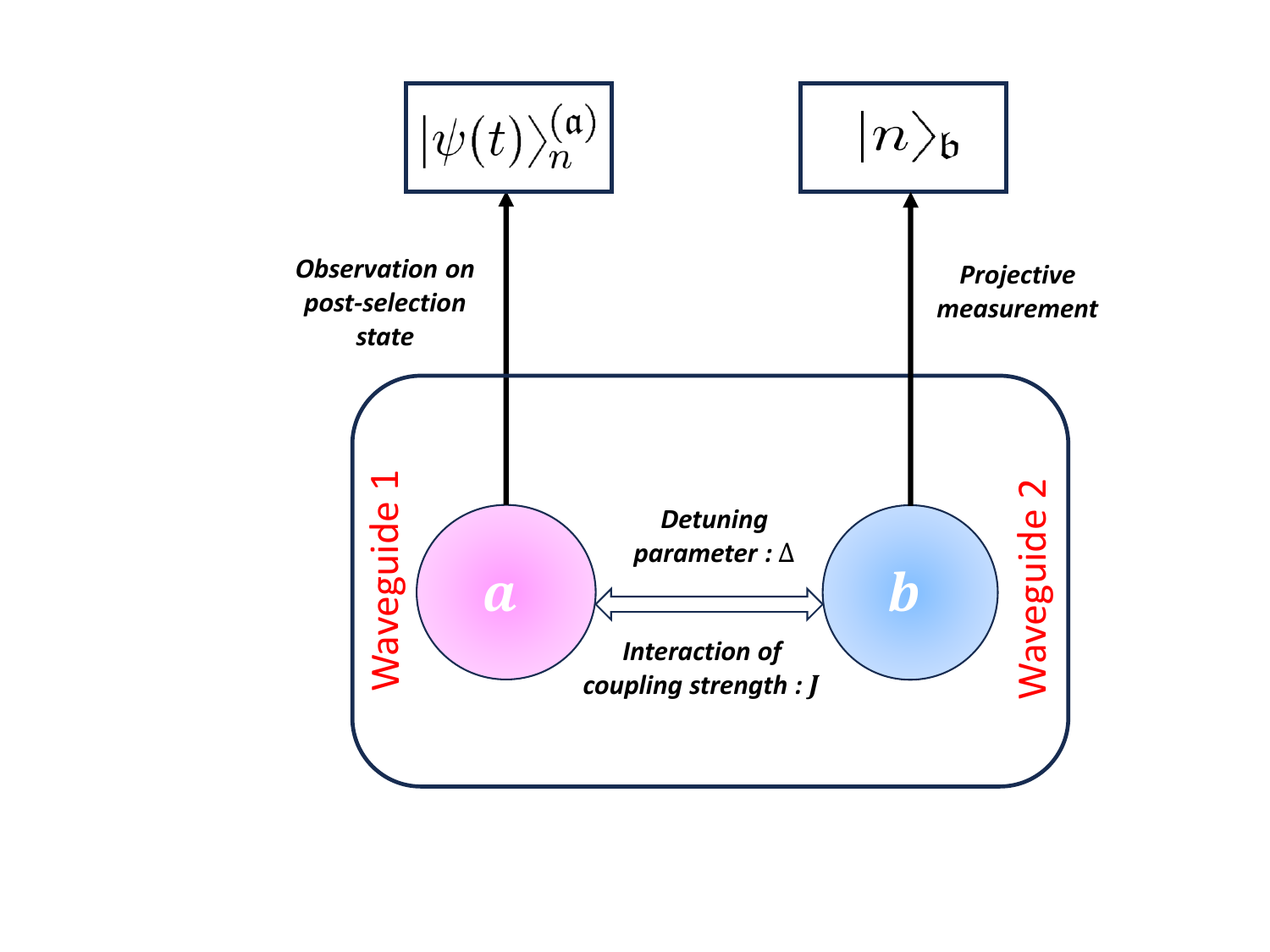}
\caption{Schematic diagram of linked waveguide system where  $\Delta$ reflects the energy detuning parameter between the modes, and the coefficient $J$ describes the tunneling rate of photons.}
\label{schematic_diagram}
\end{center}
\end{figure}

With the above picture in mind, we, in the present work,  investigate the generation of a set of nonclassical states realized via conditional measurement process on a photonic degree of freedom in a linked waveguide system.  Starting with a
bimodal squeezed vacuum input state, we, following an interaction period, projectively measure $n$-th Fock state in the output of one subsystem.  Nonclassical Hermite polynomial excited states, endowed with \textit{parameters  dependent on the extent of time as well as the strength of the interaction between the modes}, are generated in the complementary subsystem.  Hermite polynomial states has been investigated earlier [\cite{BHY1991}].  To study the said nonclassicality of the
post-measurement states obtained here, we compute the corresponding phase space Wigner distribution [\cite{A2013}], the Mandel parameter [\cite{M1979}] and the variance of the quadrature operator. A schematic diagram of the procedure considered here is given in Fig. \ref{schematic_diagram}. 

\section{Generating post-measurement states}
\label{state}
\setcounter{equation}{0}
Here we consider two coupled Bosonic modes described by  their canonical operators $(a, a^{\dagger})$ and $(b, b^{\dagger})$, respectively: $[a, a^{\dagger}]=1, [b, b^{\dagger}]=1$. The corresponding number operators read $\mathfrak{n_{a}} \equiv a^{\dagger} a,  \mathfrak{n_{b}} \equiv b^{\dagger}b$. The pertinent Hamiltonian demonstrating the interaction of the fields, say in a waveguide beamsplitter, is assumed [\cite{LA2013}] to be
\beq
\mathrm{H} = \Delta (a^{\dagger}a-b^{\dagger}b) +J (a b^{\dagger}+ a^{\dagger}b),
\label{hamiltonian}
\eeq
where $\Delta$ characterizes the detuning between two modes, and the coefficient $J$ represents the hopping rate of photons produced via the overlap between the normal modes realized in the waveguide. The well-known two mode construction of $su(1,1)$ Lie algebra reads as
\beq
 [K_{0}, K_{\pm}] = \pm\, K_{\pm}, [K_{+}, K_{-}] = - 2 K_{0}, \quad K_{+}= a^{\dagger} b^{\dagger}, K_{-}= a b, 
K_{0}= \tfrac{1}{2} \big(\mathfrak{n_{a}} + \mathfrak{n_{b}} + 1\big).
\label{su(1,1)}
\eeq
For subsequent use we quote the Baker-Campbell-Hausdorff decomposition formula for the  $su(1,1)$ group element [\cite{T1985}]:
\beq
\exp\lg-\big(\xi K_{+} - \xi^{*} K_{-}\big)\rg = \exp \lg - e^{i \phi} \tanh r\; K_{+}\rg 
\frac{1}{(\cosh r)^{2\, K_{0}}}\;\exp \lg  e^{-i \phi} \tanh r\; K_{-}\rg,
\label{K-decomp}
\eeq
where the complex  coefficient $\xi = r\, \exp(i \phi) $ specifies an arbitrary $su(1,1)$ element.

\par

At $t=0$ we start with a squeezed two mode vacuum state in the coupled waveguide:
\beq
|\Psi(0)\rangle = \exp\lg-\big(\xi a^{\dagger} b^{\dagger} - \xi^{*} a b \big)\rg 
\big|0, 0\rangle =  \tfrac{1}{\cosh r}\,\exp \lg - e^{i \phi} \tanh r\; a^{\dagger} b^{\dagger}\rg
\big|0, 0\rangle,
\label{t0-state}
\eeq
where the factorization rule (\ref{K-decomp}) has been employed.
The unitary evolution of the two mode state (\ref{t0-state}) for an interaction time $t$ reads
\beq
|\Psi(t)\rangle = \mathrm{U}(t) |\Psi(0)\rangle, \quad 
\mathrm{U}(t) = \exp(-i \mathrm{H}t),
\label{unitary-evol}
\eeq
and may be constructed  [\cite{LA2013}] by utilizing operators in the Heisenberg picture:
\bea
\begin{pmatrix}
	a(-t)\\b(-t)
\end{pmatrix}&=&\mathrm{U}(t)\,\begin{pmatrix}
	a\\b
\end{pmatrix}\, \mathrm{U}(t)^{\dagger}\nn\\
&=&\begin{pmatrix}
\sqrt{1- \mu^{2} \sin^{2} \widehat{J} t} \,\exp(i \varphi) &
i \mu \sin \widehat{J} t\\
i \mu \sin \widehat{J} t & 
\sqrt{1- \mu^{2} \sin^{2} \widehat{J} t} \,\exp(- i \varphi)
\end{pmatrix}
\begin{pmatrix}
	a\\b
\end{pmatrix},
\label{evolution}
\eea
where the parameters are given by $\widehat{J}=\sqrt{J^{2}+\Delta^{2}}, \delta= \Delta/\widehat{J}, \mu= J/\widehat{J}, \tan \varphi = \delta \tan \widehat{J} t$. The time evolution of the bipartite state (\ref{unitary-evol}) now assumes the form
\beq
|\Psi(t)\rangle =  \tfrac{1}{\cosh r}\,\exp \lg - e^{i \phi} \tanh r\; a^{\dagger}(-t) b^{\dagger}(-t)\rg
\big|0, 0\rangle.
\label{state-t}
\eeq
The  unitary matrix (\ref{evolution}) recasts the two mode state (\ref{state-t}) via an exponentiation of quadratic operators:
\beq
|\Psi(t)\rangle =  \tfrac{1}{\cosh r}\,\exp \lg \Omega \, a^{\dagger\;2}\rg\;
\exp \lg 2\, (- \omega a^{\dagger})\,(\sigma b^{\dagger}) - (\sigma b^{\dagger})^{2}\rg 
\big|0, 0\rangle,
\label{st-t-Hermite}
\eeq
where the time dependent coefficients are listed as follows:
\bea 
\Omega  &=& \exp\lg i(\phi - \varphi +\tfrac{\pi}{2})\rg\, \mu \; \tanh r\; \sin \widehat{J} t\;
\sqrt{1- \mu^{2} \sin^{2} \widehat{J} t},\nn\\
\omega  &=& \exp\lg \tfrac{i}{2}(\phi - \varphi +\tfrac{\pi}{2})\rg 
\tfrac{\sqrt{\tanh r}\, \big(1- 2 \mu^{2} \sin^{2} \widehat{J} t\big)}
{2\,\sqrt{\mu \sin \widehat{J} t} \,\big(1-  \mu^{2} \sin^{2} \widehat{J} t\big)^{\frac{1}{4}}},
\nn\\
\sigma  &=&  \exp\lg \tfrac{i}{2}(\phi + \varphi - \tfrac{\pi}{2})\rg \, 
\sqrt{\mu \; \tanh r\; \sin \widehat{J} t}\;
\big(1-  \mu^{2} \sin^{2} \widehat{J} t\big)^{\frac{1}{4}}
\label{st-variables}.
\eea
Explicit representation of time dependence in the l.h.s. of (\ref{st-variables}) is,  however, omitted for maintaining  simplicity in notation.
The above complex coefficients maintain a real proportionality structure 
\beq
\Omega = \zeta \, \omega^{2}, \qquad \zeta = \tfrac{4\,\mu^{2} \sin^{2} \widehat{J} t\; \big(1-  \mu^{2} \sin^{2} \widehat{J} t\big)}
{\big(1- 2\, \mu^{2} \sin^{2} \widehat{J} t\big)^{2}} \in \mathbb{R}
\label{zeta-real}
\eeq
that we intend to apply later. We also note that in the near-null detuning regime $J \gg \Delta$, the limiting values of the parameters are given by $\widehat{J} \approx J, \mu \approx 1, \varphi \approx 0$, and, consequently, it follows:
\beq 
\Omega  \approx \exp\lg i(\phi  +\tfrac{\pi}{2})\rg\, \tfrac{1}{2} \tanh r\; \sin (2\widehat{J} t),\quad
\omega  \approx \exp\lg \tfrac{i}{2}(\phi +\tfrac{\pi}{2})\rg 
\tfrac{\sqrt{\tanh r}\, \cos(2\widehat{J} t)}
{\sqrt{2\, \sin(2 \widehat{J} t})}.
\label{ultra-par}
\eeq

\par

The Gaussian operator construction of the bimodal state (\ref{st-t-Hermite}) facilitates representation of projective measurement in one mode. Proceeding towards this we quote  the 
generating function property [\cite{R1971}] of the Hermite polynomials:
\beq
\exp\lg 2 \mathcal{X} \vartheta - \vartheta^{2}\rg =\sum_{\ell = 0}^{\infty} H_{\ell}\big(\mathcal{X}\big)
\tfrac{\vartheta^{\ell}}{\ell!}, \quad H_{\ell}\big(- \mathcal{X}\big) =(-1)^{\ell}\, H_{\ell}\big(\mathcal{X}\big).
\label{Hermite-sum}
\eeq
The matrix element of the Fock state of the B-mode ${}_{\mathfrak{b}}\langle n|b^{\dagger \;\ell}|0, 0\rangle = \delta_{n, \ell}\;\sqrt{\ell !}\;  |0\rangle_{\mathfrak{a}}$ now allows us to compute its overlap with the bipartite state (\ref{st-t-Hermite}): 
\beq
{}_{\mathfrak{b}}\langle n|\Psi(t)\rangle = (-1)^{n}\, \tfrac{\sigma^{n}}{\sqrt{n!}\;\cosh r}
\exp\lg \Omega a^{\dagger\;2}\rg  H_{n}\big(\omega a^{\dagger} \big) |0\rangle_{\mathfrak{a}},
\label{project-st}
\eeq
where the squared norm of the projection reads
\beq
\left|{}_{\mathfrak{b}}\langle n|\Psi(t)\rangle\right|^{2} = 
\tfrac{|\sigma|^{2\,n}}{n!\;\cosh^{2} r}\;  {}_{\mathfrak{a}}\langle 0|  H_{n}\big(\omega^{*} a\big)\, 
\exp\lg \Omega^{*} a^{2}\rg 
\exp\lg \Omega a^{\dagger\;2}\rg  H_{n}\big(\omega a^{\dagger} \big) |0\rangle_{\mathfrak{a}}.
\label{st-norm}
\eeq
In the present context the time $t$ refers to the period elapsed before the projective measurement is executed on the B-mode.

\par

To assist the evaluation of squared norm (\ref{st-norm}) we utilize the decomposition of unity in the basis of coherent states [\cite{A2013}]:
\beq
|\alpha\rangle \equiv \mathrm{D}(\alpha) |0\rangle =  \exp\lg \tfrac{-|\alpha|^{2}}{2}\rg \sum_{n=0}^{\infty} \tfrac{\alpha^{n}}{\sqrt{n!}} |n\rangle,
\;\; \mathrm{D}(\alpha) = \exp\big(\alpha a^{\dagger}- \alpha^{*} a\big), \;\;
\int \tfrac{\mathrm{d}^{2} \alpha}{\pi} |\alpha \rangle\, \langle \alpha | =\mathbb{I}. 
\label{coherent}
\eeq
The matrix element  quoted in (\ref{st-norm}) now admits computation via an integration over the phase space. As it is required for our subsequent use, we  consider up to the second moment of the integral kernel: 
\bea
\mathfrak{I}_{n}^{(k)} &\equiv&  \langle 0|  H_{n}\big(\omega^{*} a\big)\, 
 \exp\lg \Omega^{*} a^{2}\rg a^{k}a^{\dagger\,k} 
 \exp\lg \Omega a^{\dagger\;2}\rg  H_{n}\big(\omega a^{\dagger} \big) |0\rangle \nn\\
&=& \int \tfrac{\mathrm{d}^{2} \alpha}{\pi} \, \exp \lg  - |\alpha|^{2} + \Omega^{*} \alpha^{2} 
+ \Omega \alpha^{*\,2} \rg |\alpha|^{2 k}  H_{n}\big(\omega^{*} \alpha\big)\,
 H_{n}\big(\omega  \alpha^{*}\big),
\label{I-n-int}
\eea
where $ k \in \{0, 1, 2\}$. We note that the matrix element  on the right hand side in (\ref{st-norm})  corresponds to $ \mathfrak{I}_{n}^{(0)}$. The convergence of the Gaussian integral (\ref{I-n-int}) imposes a restriction on the parametric space: 
$|\Omega| <\tfrac{1}{2}\,\Rightarrow \mu \, \tanh r < \tfrac{1}{2}$. Now we specify  our recipe for evaluating the phase space integral (\ref{I-n-int})  by exploiting the bilocal generating function of the Hermite polynomials [\cite{R1971}]:
\beq
\sum_{n = 0}^{\infty}  H_{n}\big(\mathcal{X}\big)\, H_{n}\big(\mathcal{Y}\big)\;\tfrac{\tau^{n}}{n!}
= \tfrac{1}{\sqrt{1 - 4 \tau^{2}}}\;\exp \lg \tfrac{4}{1 - 4 \tau^{2}} \ls \mathcal{X}\, \mathcal{Y}\;
	\tau - \lo \mathcal{X}^{2} + \mathcal{Y}^{2} \rc \tau^{2} \rs \rg,
\label{bilocal}
\eeq
where $\tau$ is an arbitrary expansion variable.    
Towards procuring the integrals $\mathfrak{ I}^{(k)}_{n}$ we introduce  a formal sum 
\beq
\mathfrak{I}^{(k)}(\tau)  \equiv \!\!\sum_{n = 0}^{\infty}\mathfrak{ I}^{(k)}_{n}\tfrac{\tau^{n}}{n!} = \!\! \int \tfrac{\mathrm{d}^{2} \alpha}{\pi}  \exp \lg  - |\alpha|^{2} + \Omega^{*} \alpha^{2} 
+ \Omega \alpha^{*\,2} \rg |\alpha|^{2 k}\!\! \sum_{n=0}^{\infty}  H_{n}\big(\omega^{*} \alpha\big)
H_{n}\big(\omega  \alpha^{*}\big) \tfrac{\tau^{n}}{n!},
\label{sum-I-k}
\eeq
 and, following the implementation of  (\ref{bilocal}), explicitly compute it via Gaussian integration. For the  present purpose of deriving the normalization of states  
(\ref{project-st}, \ref {st-norm}) it suffices to consider $\mathfrak{I}^{(0)}(\tau)$.  As $\zeta$ given in (\ref{zeta-real}) is real, the sum  $\mathfrak{I}^{(0)}(\tau)$ assumes a factorized form:
\beq
\mathfrak{I}^{(0)}(\tau) = \tfrac{1}{\sqrt{\mathfrak{D}_{+}^{\tau}\, \mathfrak{D}_{-}^{\tau}}},\quad \mathfrak{D}_{\pm}^{\tau} = 1 \mp 2 \zeta |\omega|^{2} \pm 2 \tau \big(1 \mp 2 \zeta |\omega|^{2} 
\mp 2 |\omega|^{2}\big).
\label{I-factor}
\eeq
Recasting the series sum (\ref{I-factor}) via  binomial theorem
\bea
\mathfrak{I}^{(0)} (\tau)& =& \tfrac{1}{\sqrt{\Theta}}\;\sum_{n = 0}^{\infty} 2^{n} \lg \sum_{k = 0}^{n} (-1)^{k} \binom {n}{k}
\lo \tfrac{1}{2} \rc_{k}\lo \tfrac{1}{2} \rc_{n-k} \lo 1 - \tfrac{2\, |\omega|^{2}}{1-2\,
	 \zeta \, |\omega|^{2} }\rc^{k} \times \right.\nn\\  
 &&\times \left.\lo 1 + \tfrac{2\, |\omega|^{2}}{1+2\,
	 \zeta \, |\omega|^{2} }\rc^{n-k}\rg\; \tfrac{\tau^{n}}{n!},
\label{I-expn}
\eea
we extract the positive definite squared norm as follows:
\bea
\mathfrak{I}_{n}^{(0)} &=& \tfrac{2^{n}}{\sqrt{\Theta}} \sum_{k = 0}^{n} (-1)^{k} \binom {n}{k}
\lo \tfrac{1}{2} \rc_{k}\lo \tfrac{1}{2} \rc_{n-k} \lo 1 - \tfrac{2\, |\omega|^{2}}{1-2\,
	\zeta \, |\omega|^{2} }\rc^{k}   
\lo 1 + \tfrac{2\, |\omega|^{2}}{1+2\,
	\zeta \, |\omega|^{2} }\rc^{n-k}\nn\\
&\equiv& \tfrac{2^{n}}{\sqrt{\Theta}}\; \mathrm{P}_{n} \lg \tfrac{1}{2}, \tfrac{1}{2};  \mathsf{x}_{1}, \mathsf{x}_{2}\rg,
\label{In-expn}
\eea
where $\mathsf{x}_{1} = -1+  \tfrac{2\, |\omega|^{2}}{1-2\,\zeta \, |\omega|^{2}}, \mathsf{x}_{2} = 1+  \tfrac{2\, |\omega|^{2}}{1+2\,\zeta \, |\omega|^{2}}, \Theta = 1 - 4 \zeta^{2} \, |\omega|^{4}$.  In (\ref{I-expn}) we have used the notation $(\mathsf{a})_{0}=1, (\mathsf{a})_{n} = \mathsf{a} (\mathsf{a}+1)\ldots (\mathsf{a}+n-1)$. The bivariate polynomial $\mathrm{P}_{n} \lg \mathsf{a}_{1}, \mathsf{a}_{2}; \mathsf{z}_{1}, \mathsf{z}_{2}\rg $ in (\ref{In-expn}) denotes the coefficient in  expansion of the product structure
\bea
\lo 1- \mathsf{z}_{1} \tau\rc^{\mathsf{-a}_{1}} \;\lo 1- \mathsf{z}_{2} \tau\rc^{\mathsf{-a}_{2}} 
&=& \sum_{n = 0}^{\infty} \mathrm{P}_{n} \lg \mathsf{a}_{1}, \mathsf{a}_{2}; \mathsf{z}_{1}, \mathsf{z}_{2}\rg \tfrac{\tau^{n}}{n!},\nn\\
\mathrm{P}_{n} \lg \mathsf{a}_{1}, \mathsf{a}_{2}; \mathsf{z}_{1}, \mathsf{z}_{2}\rg &=& 
\sum_{k = 0}^{n} \binom{n}{k} \lo \mathsf{a}_{1} \rc_{k}
\lo \mathsf{a}_{2} \rc_{n-k}\; \mathsf{z}_{1}^{k}\; \mathsf{z}_{2}^{n-k}.
\label{prod-pol}
\eea
The polynomial $\mathrm{P}_{n} \lg \mathsf{a}_{1}, \mathsf{a}_{2}; \mathsf{z}_{1}, \mathsf{z}_{2}\rg$  is an element of the general class of hypergeometric product polynomials investigated in [\cite{K2019}].
The construction (\ref{In-expn}) now provides the norm $\mathfrak{I}_{n}^{(0)}$ for the post-measurement states with  arbitrary $n$. We reproduce first few elements herein:
\bea
\mathfrak{I}_{0}^{(0)} &=& \tfrac{1}{\sqrt{\Theta}},\;\; 
\mathfrak{I}_{1}^{(0)} = \tfrac{4\,|\omega|^{2}}{\Theta^{3/2}},\;\;
\mathfrak{I}_{2}^{(0)} = \tfrac{4}{\sqrt{\Theta}} \lg 
1- \tfrac{4\, |\omega|^{4}}{\Theta} 
- \tfrac{8\, \zeta  |\omega|^{4}}{\Theta} 
+ \tfrac{12\, |\omega|^{4}}{\Theta^{2}} \rg, \nn\\
\mathfrak{I}_{3}^{(0)} &=& \tfrac{48 \, |\omega|^{2}}{\Theta^{3/2}} 
\lg 3 - \tfrac{12 \,|\omega|^{4}}{\Theta} - \tfrac{24\, \zeta |\omega|^{4}}{\Theta} + 
\tfrac{20\, |\omega|^{4}}{\Theta^{2}} \rg.
\label{I123}
\eea
As a consistency check on the above derivation of the projection element (\ref{project-st}) we demonstrate the validity of the unitarity constraint. Employing the phase space representation (\ref{I-n-int}) the sum of  squared projectors assumes the form
\bea
\sum_{n=0}^{\infty} \left|{}_{\mathfrak{b}}\langle n|\Psi(t)\rangle\right|^{2} &=&
\tfrac{1}{\cosh^{2} r}\,\int \tfrac{\mathrm{d}^{2} \alpha}{\pi} \, \exp \lg  - |\alpha|^{2} + \Omega^{*} \alpha^{2} 
+ \Omega \alpha^{*\,2} \rg  \times \nn\\
&&\times \sum_{n=0}^{\infty} H_{n}\big(\omega^{*} \alpha\big) 
 H_{n}\big(\omega  \alpha^{*}\big)
\tfrac{|\sigma|^{2\,n}}{n!}.
\label{unitary-prf}
\eea
Computing the series on the right hand side via application of the  generating function (\ref{bilocal}), and subsequent  Gaussian integration on the phase space, we obtain the desired sum rule:
\beq
\sum_{n=0}^{\infty} \left|{}_{\mathfrak{b}}\langle n|\Psi(t)\rangle\right|^{2} = 
\tfrac{1}{ \cosh^{2} r \,\sqrt{\mathfrak{D}_{+}^{|\sigma|^{2}}\, \mathfrak{D}_{-}^{|\sigma|^{2}}}}
  = 1,
\label{unitary}
\eeq
where in the second equality we applied the explicit values of the parameters listed in 
(\ref{st-variables}). Combining the results (\ref{project-st}, \ref{st-norm}, \ref{In-expn}) we now reproduce 
the normalized  state for the residual  A degree of freedom generated as a consequence of the  projective measurement on its B counterpart:
\beq
|\psi(t)\rangle^{(\mathfrak{a})}_{n} = \tfrac{{}_{\mathfrak{b}}\langle n|\Psi(t)\rangle}
{\left|{}_{\mathfrak{b}}\langle n|\Psi(t)\rangle\right|} 
= \tfrac{1}{\sqrt{\mathfrak{I}_{n}^{(0)}}}\,\exp\lg \Omega a^{\dagger\;2}\rg  H_{n}\big(\omega a^{\dagger} \big) |0\rangle_{\mathfrak{a}}.
\label{n-state}
\eeq
In  (\ref{n-state}) we have omitted an overall phase. Periodicity  of the coefficients (\ref{st-variables}) imparts a cyclic structure on  the state (\ref{n-state})  where the period in scaled time $\widehat{J} t$ equals  $ \pi$. For a near-null detuning parameter 
($J \gg\Delta, \mu \approx 1$), however, a new approximate periodicity emerges at the following scaled time: $\widehat{J} t|_{\mathrm{period}} \approx \tfrac{\pi}{2}$.
It directly follows from the observed limiting values  listed in (\ref{ultra-par}). Following  our construction (\ref{st-variables}) we note that  the limiting values of the parameters $\Omega \sim t, \;\omega  \sim \tfrac{1}{\sqrt{t}}, \sigma \sim \sqrt{t}, \zeta \rightarrow t^{2}$ hold as $t \rightarrow 0$. In this limit the 
composition of the post-measurement state  (\ref{In-expn}, \ref{n-state}) allows us to infer 
\beq
\left.|\psi(t)\rangle^{(\mathfrak{a})}_{n}\right|_{t \rightarrow 0} \sim \Big( \tfrac{\omega}{|\omega|} \Big)^{n}  a^{\dagger\;n}\,
|0\rangle_{\mathfrak{a}} \sim |n\rangle_{\mathfrak{a}}.
\label{st-t-0}
\eeq
Our construction maintains the boundary condition that in the $t \rightarrow 0$ limit the initial bipartite state (\ref{t0-state}), following its projection on the Fock state of the B-mode 
$|n\rangle_{\mathfrak{b}}$, produces the corresponding Fock state of the residual A-mode $|n\rangle_{\mathfrak{a}}$.
The class of states given in  (\ref{n-state}) has been previously obtained in  [\cite{BHY1991}]. In our case it materializes due to a projective measurement  exercised on a degree of freedom of an interacting system, and, consequently, the parameters 
(\ref{st-variables}) characterizing the state depend on time of interaction between the modes.
The power series expansion of the Hermite polynomials [\cite{R1971}] implies that the quantum state outlined in (\ref{n-state}) embodies superposition of squeezed number states. This feature provides the underlying basis  of nonclassical characteristics described in the sequel. Moreover, the parity property of the Hermite polynomials splits the ensemble of  states (\ref{n-state}) in an even and an odd parity sectors:
\beq
\Pi_{\mathfrak{a}}\;  H_{n}\;\big(\omega a^{\dagger} \big) \Pi_{\mathfrak{a}} =  H_{n}\big(-\omega a^{\dagger} \big) = (-1)^{n}  H_{n}\big(\omega a^{\dagger} \big) \; \Rightarrow \; 
\Pi_{\mathfrak{a}} |\psi(t)\rangle^{(\mathfrak{a})}_{n} = (-1)^{n}\;|\psi(t)\rangle^{(\mathfrak{a})}_{n},
\label{parity}
\eeq
where the parity operator for the A-mode is given by $\Pi_{\mathfrak{a}} = \exp(i \pi a^{\dagger} a)$. The above construction immediately establishes the orthogonality of the even and odd parity states:
\beq
{}^{\quad(\mathfrak{a})}_{2n+1}\langle \psi(t)|\psi(t)\rangle^{(\mathfrak{a})}_{2n} = 0.
\label{even-odd}
\eeq

\section{Wigner distribution of the post-measurement states}
\label{wigner}
\setcounter{equation}{0}
In this section we focus on the Wigner quasiprobability distribution [\cite{A2013}] for the  states (\ref{n-state}) realized via projective measurement on the B-mode. As  we  presently scrutinize the state  of the  A degree of freedom in the post-measurement scenario, we henceforth omit the index referring to the said mode. We start with the 
 $\mathrm{P}$-representation of the density matrix that may be formally constructed for any state [\cite{S1963}, \cite{G1963}]. In the overcomplete coherent state basis the density matrix may be expressed as a diagonal sum:
\beq
\rho = \int \mathrm{P}(\alpha, \alpha^{*})\,|\alpha\rangle \,\langle \alpha|\, 
\mathrm{d}^{2} \alpha, \;\; \int \mathrm{P}(\alpha, \alpha^{*})\, \mathrm{d}^{2} \alpha = 1.
\label{P_def_S}
\eeq
Using Fourier transformation on a complex plane a useful method of evaluation for the   
$\mathrm{P}$-representation has been established [\cite{M1967}]:
\beq
\mathrm{P}(\alpha, \alpha^{*})=\tfrac{\exp(|\alpha|^{2})}{\pi^{2}} \,\int \langle-\beta|\rho|\beta\rangle\, \exp\big(|\beta|^{2}\big)\,\exp(\alpha \beta^{*} - \alpha^{*} 
\beta) \, \mathrm{d}^{2}\beta.
\label{P-invert_S}
\eeq
In the present example, however, the diagonal $\mathrm{P}$-representation is not bounded. Consequently, to investigate the nonclassicality of  states  (\ref{n-state}) we turn our attention to the Wigner distribution.  In the coherent state framework Wigner $\mathrm{W}$-distribution is introduced as the complex Fourier transform of the characteristic function [\cite{A2013}]:
\beq
\mathrm{W}(\alpha, \alpha^{*})= \tfrac{1}{\pi^{2}} \int \mathrm{Tr}[\rho\,\mathrm{D} (\beta)] 
\,\exp(\alpha \beta^{*}- \alpha^{*} \beta) \mathrm{d}^{2}\beta, \quad 
\int \mathrm{W}(\alpha, \alpha^{*}) \mathrm{d}^{2} \alpha = 1.
\label{Wigner_def_S}
\eeq
Wigner $\mathrm{W}$-distribution relates  to the diagonal  $\mathrm{P}$-representation via a  smoothing Gaussian  integral kernel [\cite{A2013}]:
\beq
\mathrm{W}(\alpha, \alpha^{*})= \tfrac{2}{\pi} \int \mathrm{P}(\beta, \beta^{*})
\,\exp\big(-2|\alpha -\beta|^{2}\big)\,\mathrm{d}^{2}\beta.
\label{P-W_S}
\eeq
Applying the construction (\ref{P-invert_S}) in the right hand side of the integral equation (\ref{P-W_S})  a  recipe for determination of the $\mathrm{W}$-distribution  in the present case may be 
established [\cite{AW1970}]:
\beq
\mathrm{W}(\alpha, \alpha^{*})=\tfrac{2}{\pi^{2}} \,\exp(2|\alpha|^{2})\,\int \langle-\beta|\rho|\beta\rangle\, \exp(2(\alpha \beta^{*} - \alpha^{*} \beta)) \, \mathrm{d}^{2}\beta.
\label{P-W-sim-exp_S}
\eeq
In the sequel we also utilize the Wigner $\mathrm{W}$-distribution  to evaluate the Hilbert-Scmidt metric [\cite{DMMW2000}]  that measures  the separation of two density matrices, say $\rho_{1}$ and $\rho_{2}$, on the Hilbert space:
\beq
\ls \mathrm{d_{HS}}(\rho_{1}, \rho_{2})\rs ^{2} \equiv \mathrm{Tr}\lg (\rho_{1} - \rho_{2})^{2}\rg =
\pi \int \ls \mathrm{W}_{1}(\alpha, \alpha^{*}) - \mathrm{W}_{2}(\alpha, \alpha^{*}) \rs^{2} \, \mathrm{d}^{2}\alpha,
\label{HSM}
\eeq
where the distributions $\mathrm{W}_{1}$ and $\mathrm{W}_{2}$ correspond to the density matrices $\rho_{1}$ and $ \rho_{2}$, respectively.

\par

For the  state listed in  (\ref{n-state}) the density matrix explicitly reads 
$\rho = |\psi(t)\rangle_{n} \,{}_{n}\langle \psi(t)|$. 
 Implementing the phase space integral 
(\ref{P-W-sim-exp_S}) we obtain the corresponding Wigner quasiprobability  distribution:
\bea
\mathrm{W}_{n}(\alpha, \alpha^{*})&=&\tfrac{2}{\pi^{2}\, \mathfrak{I}_{n}^{(0)}} \,\exp(2|\alpha|^{2})\,\int 
\exp \lg  - |\beta|^{2} + \Omega^{*} \beta^{2} 
+ \Omega \beta^{*\,2} + 2(\alpha \beta^{*} - \alpha^{*} \beta)\rg\;\; \times\nn\\ 
&& \times \;\; H_{n}\big(\omega^{*} \beta\big) 
H_{n}\big(- \omega  \beta^{*}\big)\, \mathrm{d}^{2} \beta.
\label{W-n-int}
\eea
Towards the evaluation of Wigner distribution $\mathrm{W}_{n}(\alpha, \alpha^{*})$ for an arbitrary $n$ we adopt a procedure that hinges on bilinear generating function of the Hermite polynomials
 (\ref{bilocal}). Introducing a supplementary quantity $\mathcal{W}_{n}$ that represents \textit{unnormalized} Wigner distribution $\big(\mathrm{W}_{n} \equiv 
 \tfrac{ \mathcal{W}_{n}}{\mathfrak{I}_{n}^{(0)}}\big)$ we consider a formal series: 
$\mathcal{W}(\tau) = \sum_{n}\mathcal{W}_{n}\, \frac{\tau^{n}}{n!}$.  The corresponding phase space integral is remodeled in a Gaussian form on account of the sum rule (\ref{bilocal}). Its  explicit computation reads
\bea
\mathcal{W}(\tau) &=& 
\tfrac{2\,\exp(2|\alpha|^{2})}{\pi\,\sqrt{\mathcal{D}_{+} \mathcal{D}_{-}}}\,
\exp\lg -4 \kappa_{r}^{2}\;\tfrac{|\omega|^{2}\, (1+ 2 \tau)}{\mathcal{D}_{+}}\rg
\exp\lg -4 \kappa_{\imath}^{2}\;\tfrac{|\omega|^{2}\, (1- 2 \tau)}{\mathcal{D}_{-}}\rg,\nn\\
\mathcal{D}_{\pm} &=& 1 \pm 2 \zeta |\omega|^{2} \pm 2 \tau \big(1 \pm 2 \zeta |\omega|^{2} 
\pm 2 |\omega|^{2}\big),
\label{W-sum}
\eea
where we  employ a scaled phase space variable $\kappa  \equiv \kappa_{r} + \imath \,\kappa_{\imath} = \frac{\alpha}{\omega} $, and, intending to maintain a simpler  notation, designate the spreads of the Gaussian fall-offs  therein by $\mathcal{D}_{\pm} (\equiv \mathfrak{D}_{\mp}^{-\tau}$ given previously  in (\ref{I-factor})).  Utilizing the generating function relation of the associated Laguerre polynomials [\cite{R1971}] the sum  $\mathcal{W}(\tau)$ may be expanded in a power series:
\bea
\mathcal{W}(\tau) &=&\mathcal{G}(\alpha, \alpha^{*}) \sum_{n=0}^{\infty} (-1)^{n}\,2^{n} \sum_{\ell =0}^{n}\mathsf{x}_{1}^{\ell} \; \mathsf{x}_{2}^{n-\ell}\; L^{\big(-\tfrac{1}{2}\big)}_{\ell } \lg -\tfrac{8  \kappa_{\imath}^{2}\;|\omega|^{4}}{\big(1- 2 \,\zeta |\omega|^{2}\big) \big(1- 2 \,\zeta |\omega|^{2}\ - 2 |\omega|^{2}\big)}\rg \; \times\nn\\
&&\times \;\;L^{\big(-\tfrac{1}{2}\big)}_{n-\ell }\lg \tfrac{8  \kappa_{r}^{2}\;|\omega|^{4}}{\big(1+ 2 \,\zeta |\omega|^{2}\big) \big(1+ 2 \,\zeta |\omega|^{2}\ + 2 |\omega|^{2}\big)}\rg  \tau^{n},
\label{W-gen-fn}
\eea 
where the Gaussian prefactor  reads
\bea
\mathcal{G}(\alpha, \alpha^{*}) &\equiv& \tfrac{2 \exp (2 |\alpha|^{2})}{\pi \, \sqrt{\Theta }}\exp \lg - \tfrac{4 \kappa_{r}^{2}\;|\omega|^{2}}{1+ 2 \,\zeta |\omega|^{2}}\rg
\exp \lg - \tfrac{4 \kappa_{\imath}^{2}\;|\omega|^{2}}{1- 2 \,\zeta |\omega|^{2}}\rg\nn\\
&=& \tfrac{2}{\pi\,\sqrt{\Theta }}
\exp\lg -2 \,\tfrac {1 + 4 \zeta^{2} \, |\omega|^{4}}{\Theta }\; |\alpha|^{2} + 4 \zeta\;
\tfrac{\langle \omega,\alpha \rangle_{2} }
{\Theta}\rg.
\label{G}
\eea
In (\ref{G})  we have used the notation $ \langle \omega,\alpha \rangle_{p} \equiv \omega^{*\,p}\alpha^{p} + \omega^{p}\alpha^{*\,p},\; p\in \mathbb{Z}_{> 0}$. 
Successive terms in the Taylor expansion of $\mathcal{W}(\tau)$, and necessary insertion of corresponding normalization constants $\mathfrak{I}_{n}^{(0)}$ specified in (\ref{In-expn}), now systematically produce the Wigner distribution for the $n$-th 
post-measurement photon state  (\ref{n-state}):
\bea
\mathrm{W}_{n}(\alpha, \alpha^{*}) &=& (-1)^{n}\,2^{n} n!\;\mathcal{G}(\alpha, \alpha^{*}) \tfrac{1}{\mathfrak{I}_{n}^{(0)}} \sum_{\ell =0}^{n}\mathsf{x}_{1}^{\ell} \; \mathsf{x}_{2}^{n-\ell}\; L^{\big(-\tfrac{1}{2}\big)}_{\ell } \lg -\tfrac{8  \kappa_{\imath}^{2}\;|\omega|^{4}}{\big(1- 2 \,\zeta |\omega|^{2}\big) \big(1- 2 \,\zeta |\omega|^{2}\ - 2 |\omega|^{2}\big)}\rg \; \times\nn\\
&&\times \;\; L^{\big(-\tfrac{1}{2}\big)}_{n-\ell }\lg \tfrac{8  \kappa_{r}^{2}\;|\omega|^{4}}{\big(1+ 2 \,\zeta |\omega|^{2}\big) \big(1+ 2 \,\zeta |\omega|^{2}\ + 2 |\omega|^{2}\big)}\rg.
\label{W-n-gen}
\eea
The Gaussian prefactor (\ref{G}) present in the $\mathrm{W}$-distribution for an arbitrary $n$-th  state listed above describe the exponential fall-off realized therein.   The said 
$\mathrm{W}$-distributions satisfy the normalization restriction (\ref{Wigner_def_S}). The first few entries expressed via phase space variables are  reproduced in Appendix A.

\par

As an indicator of nonclassicality of the post-measurement states (\ref{n-state}) we study  the negative volume of the Wigner $\mathrm{W}$-distribution [\cite{KZ2004}] defined as
\beq
\delta_{\mathrm{W}} = \tfrac{1}{2} \lg \int |\mathrm{W}(\alpha, \alpha^{*})| \mathrm{d}^{2} \alpha - 1\rg.
\label{W-negative}
\eeq
In particular,  the negative volume $\delta_{\mathrm{W}}$ measures the departure of the  $\mathrm{W}$-distribution from a classical positive definite probability density function, and its time variation reveals appearance of  states of high nonclassicality heralded via measurement  at appropriate instant.

\par
\begin{figure}
\begin{center}
\captionsetup[subfigure]{labelfont={sf}}
\subfloat[]{\includegraphics[scale=0.36,trim= 0 0 0 38,clip]{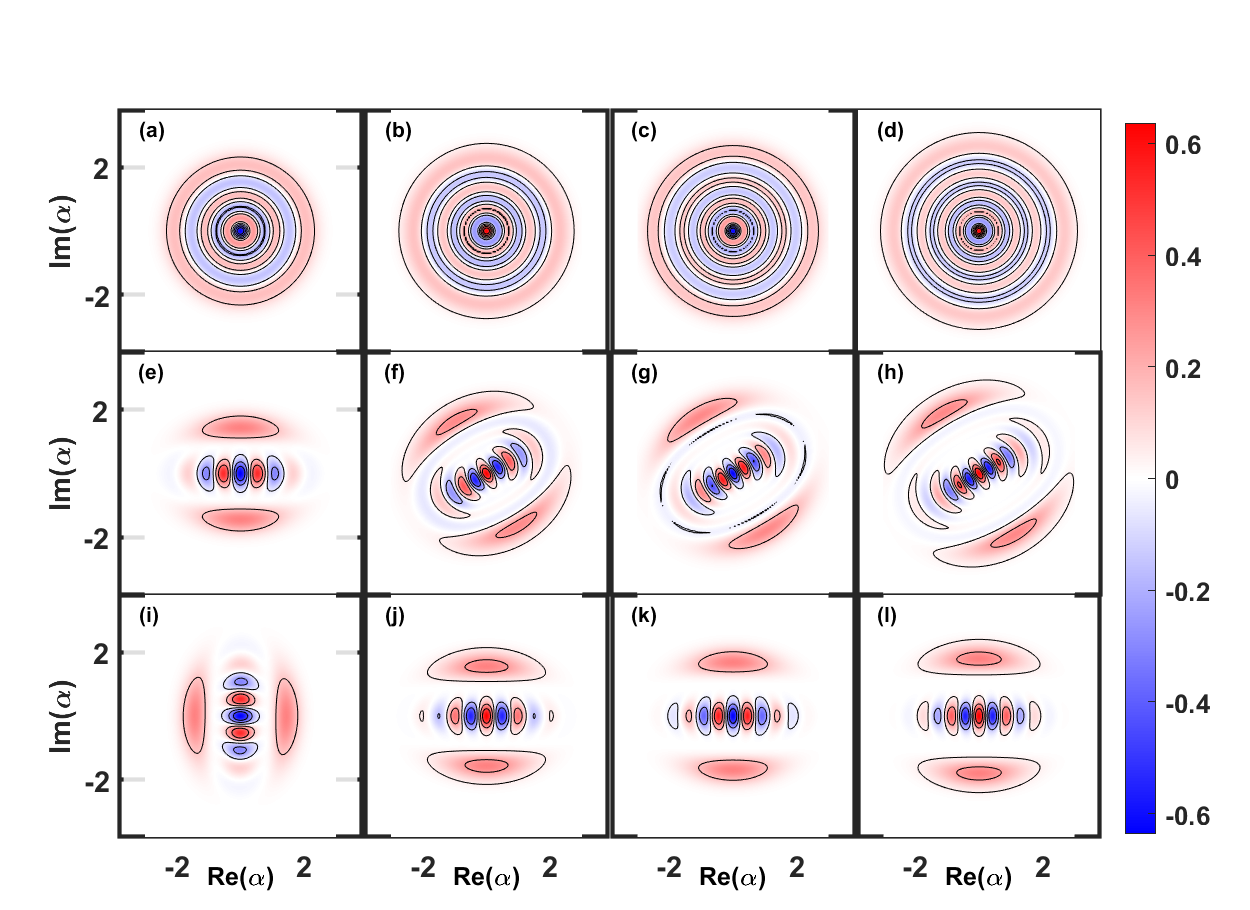}} \quad 
\subfloat[]{\includegraphics[scale=0.5,trim= 0 0 0 10,clip]{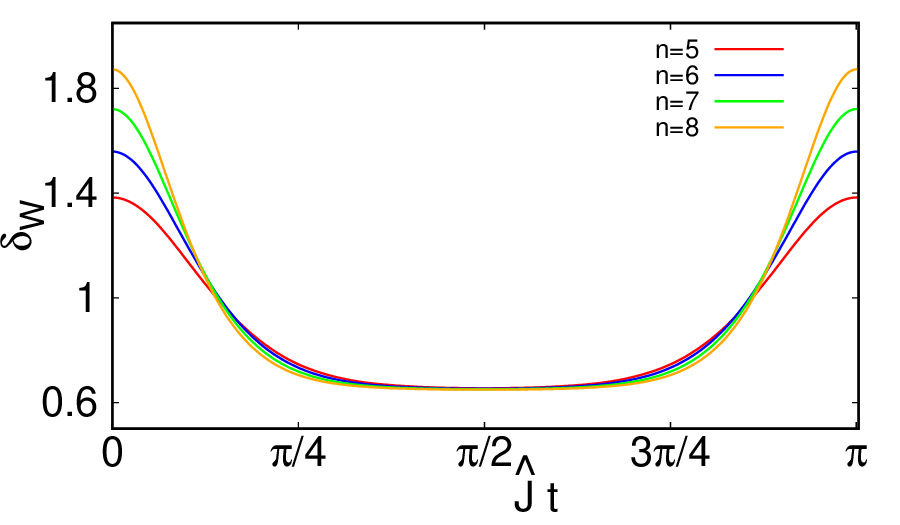}} 
\caption{(a) In the present and  subsequent diagrams we study the states $\{|\psi(t)\rangle_{n}\,| n= 5-8\}$ (unless stated otherwise) with the input squeezing parameter $\xi = 0.5$. Here we display the Wigner $\mathrm{W}$-distribution in the weak coupling domain such as  $J = 0.2\, \Delta$ at scaled interaction times $\widehat{J} t = 0, \tfrac{\pi}{6}, \tfrac{\pi}{2}$. The consecutive rows in the diagram depict the noted times in increasing order from top to bottom, whereas the columns, from left to right, mark  rising values of $n$. (b) The corresponding  Wigner negativity $\delta_{\mathrm{W}}$
is presented. In the intermediate interaction time range where the superposition of excited modes is prominent the negativity $\delta_{\mathrm{W}}$ is almost stationary and nearly equal for these values of $n$.}
\label{Wigner_j02}
\end{center}
\end{figure}
\begin{figure}
\begin{center}
\captionsetup[subfigure]{labelfont={sf}}
\subfloat[]{\includegraphics[scale=0.36,trim= 0 0 0 38,clip]{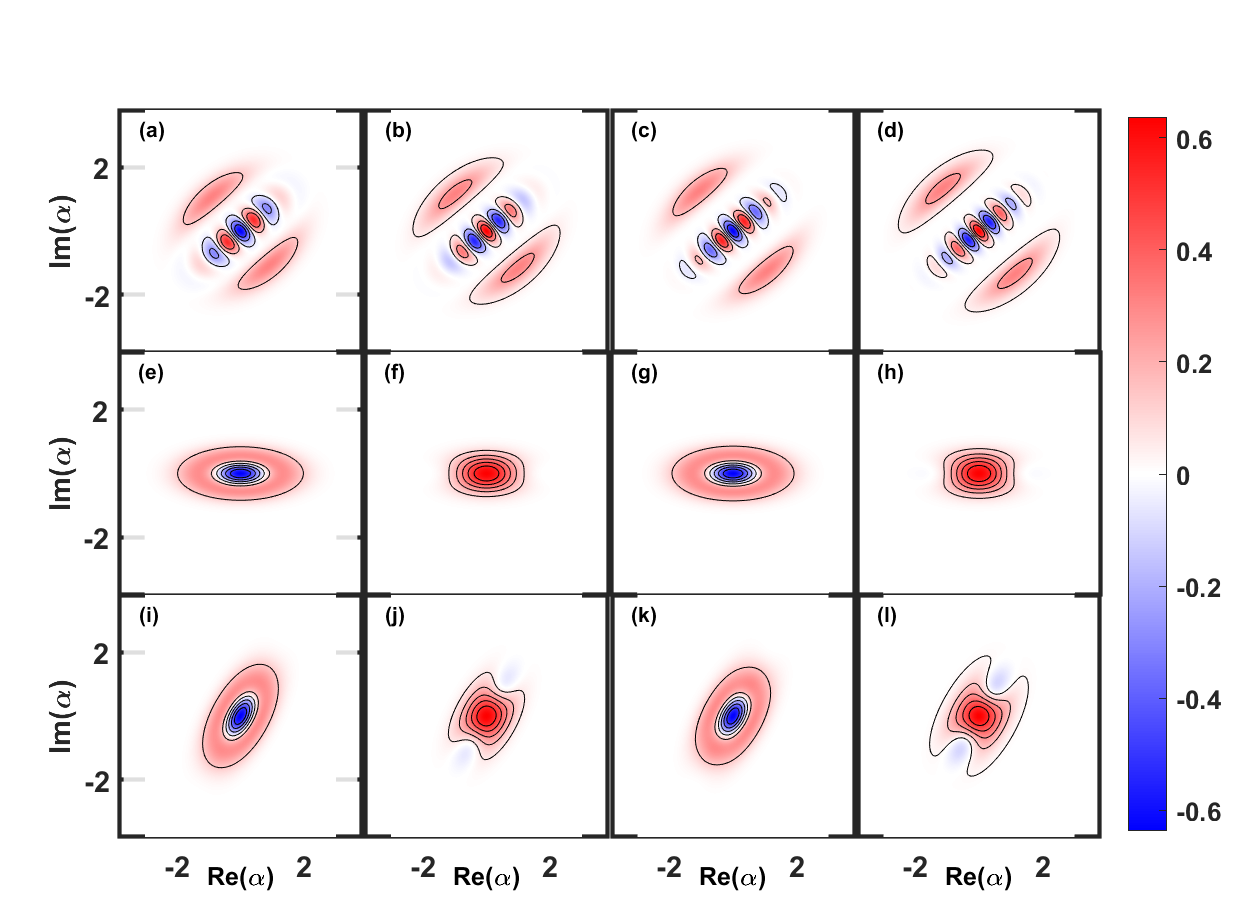}} \quad 
\subfloat[]{\includegraphics[scale=0.5,trim= 0 0 0 10,clip]{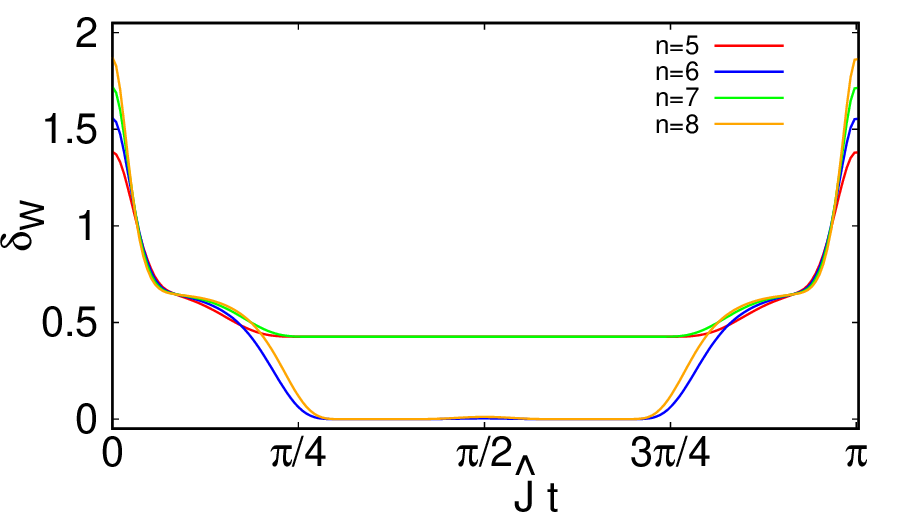}} 
\caption{ (a) We study the Wigner $\mathrm{W}$-distribution in the ultrastrong coupling realm, say  $J = 1.3\, \Delta$,  at the following  interaction periods: 
$\widehat{J} t = \tfrac{\pi}{12}, \tfrac{\pi}{2}, \tfrac{3\pi}{4}$. The row and column ordering in this diagram follows the specification given in Fig. \ref{Wigner_j02}.  Here we observe strong parity dependence in most of the interaction time range that excludes the initial period and its cyclic analogs. In particular, the even parity states therein do not exhibit significant negative values of $\mathrm{W}$-distribution, whereas the odd parity states, in contrast, demonstrate it considerably.  (b) The Wigner negativity $\delta_{\mathrm{W}}$ for the intermediate time span quantitatively confirms the preceding observation regarding  its strong parity dependence.}
\label{Wigner_j13}
\end{center}
\end{figure}

The $\mathrm{W}$-distribution and the  extent of nonclassicality of  the post-measurement state (\ref{n-state}) for various coupling coefficients are illustrated in 
Figs. \ref{Wigner_j02}, \ref{Wigner_j13}. 
For the chosen weak coupling magnitude $J = 0.2 \, \Delta$ in Fig. \ref{Wigner_j02}, we sequentially consider the said state for the quantum numbers, say, $n = 5-8$ at different measurement times. Our expressions for physical quantities relating to the state (\ref{n-state}), however, hold for an arbitrary quantum number $n$. In particular, towards visualizing the  $\mathrm{W}$-distribution for lower values of $n$ we reproduce them (Fig. \ref{Wigner_0123}) in Appendix A. 
In the first row of diagrams in Fig. \ref{Wigner_j02} (a)   that demonstrate $t \rightarrow 0$ limit of the time of interaction between the modes, we observe that the squeezed structure  of the phase space $\mathrm{W}$-distribution disappears even though our \textit{pre}-measurement initial state is the bimodal \textit{squeezed} vacuum state (\ref{t0-state}). In our construction the squeezing parameter $\Omega$ given in  (\ref{st-variables})  depends dynamically on the time of interaction while vanishing at the $t \rightarrow 0$ limit. This reflects the \textit{un}squeezed nature of the post-measurement boundary state (\ref{st-t-0}) of the A-mode. The  Wigner 
negativity $\delta_{\mathrm{W}}$ for the above cases are displayed in 
Fig. \ref{Wigner_j02} (b). In the weak coupling regime, and at smaller interaction time zone $\widehat{J}t \ll 1$ as well as its cyclic analogs,  the larger values on the quantum number $n$ for the state (\ref{n-state}) display increased  negativity $\delta_{\mathrm{W}}$ as a signature of  more prominent nonclassicality. A characteristic property, however, evident in the $\widehat{J}t \ll 1$ domain and its cyclic replicas is that the corresponding negativity changes \textit{rapidly} so that its possible implementation in experimentally obtaining a state with a specified amount of nonclassicality is likely to be challenging. In contrast, for a significant span of time in the intermediate part of the evolutionary cycle $\widehat{J}t \sim \tfrac{\pi}{2}$ the measure  $\delta_{\mathrm{W}}$ remains \textit{virtually stationary} as the corresponding interaction engenders superposition of a large number of Fock states.  This property, also shared in the ultrastrong coupling realm, augments the possibility of  controlled production and exploitation of these nonclassical states with a predetermined amount of negativity.
As a consequence of the aforesaid  superposition  the states  (\ref{n-state}) with distinct $n$ values evolve, in the $\widehat{J}t \sim \tfrac{\pi}{2}$ regime, via orbits lying  close to each other. In the ultrastrong coupling domain 
$(J \gtrsim \Delta)$, however, parity even and odd sectors operate differently. For the coupling coefficient $J = 1.3 \, \Delta$ we produce the $\mathrm{W}$-distribution and the negativity $\delta_{\mathrm{W}}$ for the states $n= 5-8$ in 
Figs. \ref{Wigner_j13} (a) and (b), respectively. For the initial time $t \rightarrow 0$ and its periodic recursions the strong and the weak coupling regimes function in a qualitatively similar way. In the intermediate time range, where the interaction between the modes is important, we find that for a wide range of interaction period  the Wigner negativity for the  even  states $n= 6, 8$ reduces to insignificantly  small values (say, $\delta_{\mathrm{W}}\big|_{n = 6} = 0.0033, \delta_{\mathrm{W}}\big|_{n = 8} = 0.0113, \widehat{J}t=\tfrac{\pi}{2} $) rendering the corresponding post-selection $\mathrm{W}$-distribution  nearly Gaussian. Employing the Hilbert-Schmidt distance $\mathrm{d_{HS}}$ between the states, given in (\ref{HSM}), this aspect is reexamined in Appendix B. Let the 
Hilbert-Schmidt distance between $|\psi(t)\rangle_{n_{1}}$ and $|\psi(t)\rangle_{n_{2}}$  states listed in (\ref{n-state}) be denoted by 
$\mathrm{d_{HS}}\lo n_{1}, n_{2}\rc$. In Fig. \ref{hs_diatance_even_odd}  we observe that in this interaction regime \textit{all} even $n$ states closely parallel the behavior of pure Gaussian $n=0$ state, where the negativity  $\delta_{\mathrm{W}}$ disappears. In contrast to the above property, the odd states  at these interaction times display substantial negativity: say, 
$\delta_{\mathrm{W}}\big|_{n = 5}  = 0.4261, \delta_{\mathrm{W}}\big|_{n = 7}  = 0.4261$ at $\widehat{J}t=\tfrac{\pi}{2}$. Specifically,  it is observed (Appendix B) that the odd $n$ states in this interaction regime lie in orbits characterized by $\mathrm{d_{HS}(1, \mathrm{odd}}) \ll 1$, and, consequently, exhibit near-identical properties. We  further study in Sec. \ref{squeeze} that the near-Gaussian even states dynamically develop other nonclassical properties such as interaction time dependent squeezing.

\begin{figure}
\begin{center}
\captionsetup[subfigure]{labelfont={sf}}
\subfloat[]{\includegraphics[scale=0.5,trim= 0 0 0 10,clip]{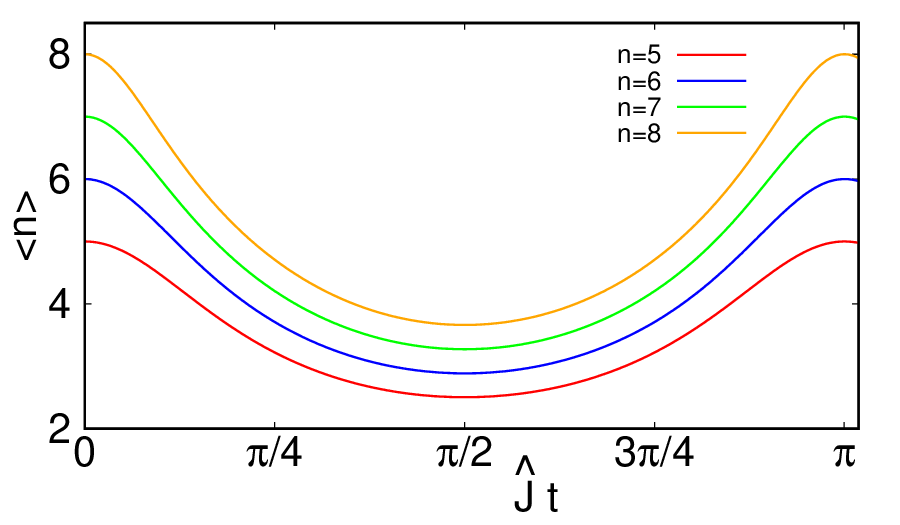}} \quad 
\subfloat[]{\includegraphics[scale=0.5,trim= 0 0 0 10,clip]{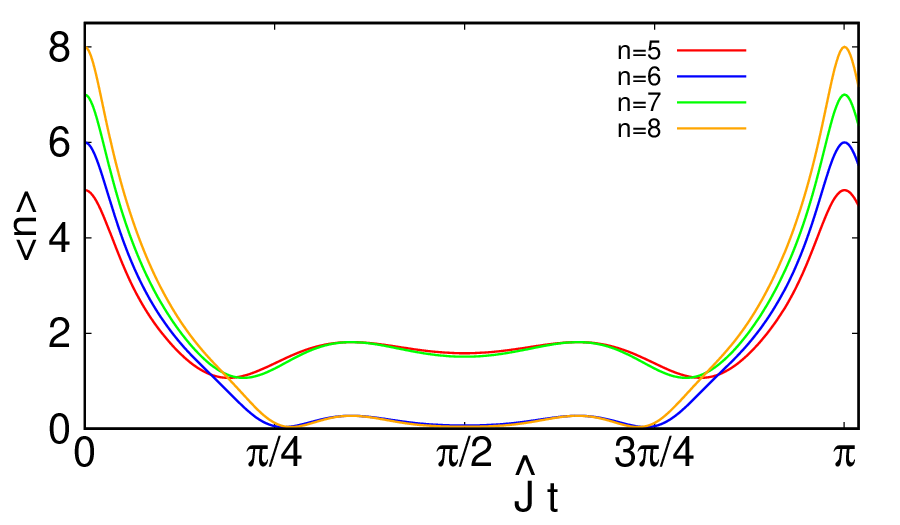}} 
\caption{We depict  the expectation value of the number operator ${}_{n}\langle \psi(t)|\mathfrak{n}|\psi(t)\rangle_{n}$ for (a) the examples of weak coupling 
$J = 0.2\, \Delta$, and (b) ultrastrong coupling $J = 1.3\, \Delta$. The $t \rightarrow 0$ limit produces qualitatively same behavior in both the  coupling domains. In the ultrastrong coupling case (b) the even and the odd parity components show distinct properties. At certain times in the intermediate interaction-dominant zone the expected number of photons for the even states reports  near-null value, whereas for the odd states thereat it describes the magnitude  ${}_{2n + 1}\langle \psi(t)|\mathfrak{n}|\psi(t)\rangle_{2n+1} \approx 1$.}
\label{expt_n1_fig}
\end{center}
\end{figure}
\section{Nonclassicality via Mandel parameter}
\label{nonclassicality}
\setcounter{equation}{0}
The Mandel parameter [\cite{M1979}] provides a measure of photon number variance, and, thereby, it captures any sub-Poissonian behavior in the photon
statistics. Consequently it  acts as  a signature of nonclassicality. For the $n$-th post-measurement state  (\ref{n-state}) it reads
\beq
Q_{n}=\tfrac{\langle (\Delta \mathfrak{n})^{2}\rangle_{n}}{{}_{n}\langle \psi(t)|\mathfrak{n} |\psi(t)\rangle_{n}}-1, \quad 
\langle (\Delta \mathfrak{n})^{2}\rangle_{n} \equiv {}_{n}\langle \psi(t)|\mathfrak{n}^{2}|\psi(t)\rangle_{n} - \lg {}_{n}\langle \psi(t)|\mathfrak{n}|\psi(t)\rangle_{n} \rg^{2},
\label{mandel-def}
\eeq
where its negative value implies sub-Poissonian statistics that signalizes antibunching of photons.
To derive the interaction time dependent Mandel parameter (\ref{mandel-def}) we employ the composition of the  state $|\psi(t)\rangle_{n}$ given in 
(\ref{n-state}), and  extract the corresponding expectation value of the photon number operator:
\beq
{}_{n}\langle \psi(t)|\mathfrak{n}|\psi(t)\rangle_{n} = \tfrac{\mathfrak{I}_{n}^{(1)}}{\mathfrak{I}_{n}^{(0)}} - 1,
\label{exp-n}
\eeq
where the coefficients on the right hand side  refer to (\ref{I-n-int}). Our recipe for obtaining these coefficients have been discussed in the context of (\ref{I-n-int}-\ref{sum-I-k}). Towards procuring the matrix element of the photon number operator (\ref{exp-n}) we first evaluate the moment $\mathfrak{ I}^{(1)} (\tau)$ given in (\ref{sum-I-k}).
These moments may be conveniently provided by admitting differentiation under the integration rule that holds for the  finite Gaussian
 integral (\ref{sum-I-k}):
\beq
\mathfrak{I}^{(k)}(\tau)   = \lim_{\mathcal{X} \rightarrow 1} \lo -\tfrac{\partial}{\partial \mathcal{X}}\rc^{k} \int \tfrac{\mathrm{d}^{2} \alpha}{\pi}  \exp \lg  - \mathcal{X} |\alpha|^{2} + \Omega^{*} \alpha^{2} 
+ \Omega \alpha^{*\,2} \rg  \sum_{n=0}^{\infty}  H_{n}\big(\omega^{*} \alpha\big)
H_{n}\big(\omega  \alpha^{*}\big) \tfrac{\tau^{n}}{n!}.
\label{diff-I-k}
\eeq
In particular, the first moment $\mathfrak{ I}^{(1)} (\tau)$  reads
\beq
\mathfrak{ I}^{(1)} (\tau) = \tfrac{1}{2\,\sqrt{\mathfrak{D}_{+}^{\tau}\, \mathfrak{D}_{-}^{\tau}}}
\lg \tfrac{1+ 2 \tau}{\mathfrak{D}_{+}^{\tau}} + \tfrac{1- 2 \tau}{\mathfrak{D}_{-}^{\tau}}\rg
\equiv \sum_{n=0}^{\infty}\mathfrak{I}_{ n}^{(1)} \;\tfrac{\tau^{n}}{n!}.
\label{I1-value}
\eeq
The equivalence relation in (\ref{I1-value})  permits systematic computation of the required matrix elements 
$\mathfrak{I}_{ n}^{(1)}$. The bivariate polynomial (\ref{prod-pol})  now facilitates explicit evaluation of the moments 
$\mathfrak{I}_{ n}^{(1)}$: 
\bea
 \mathfrak{I}_{ n}^{(1)} &=& \tfrac{2^{n-1}} {\sqrt{\Theta}} \lg \tfrac{1}{1-2 \zeta |\omega|^{2}} \ls  \mathrm{P}_{n} \lg \tfrac{3}{2}, \tfrac{1}{2};  \mathsf{x}_{1}, \mathsf{x}_{2}\rg
 + n \,\mathrm{P}_{n-1} \lg \tfrac{3}{2}, \tfrac{1}{2}; \mathsf{x}_{1}, \mathsf{x}_{2}\rg \rs \right.\nn\\
&&\left.+ \tfrac{1}{1+2 \zeta |\omega|^{2}}  \ls  \mathrm{P}_{n} \lg \tfrac{1}{2}, \tfrac{3}{2};  \mathsf{x}_{1}, \mathsf{x}_{2}\rg
 - n \,\mathrm{P}_{n-1} \lg\tfrac{1}{2}, \tfrac{3}{2}; \mathsf{x}_{1}, \mathsf{x}_{2}\rg \rs \rg.
\label{I1n-explicit}
\eea
The first few entries are listed below:
\bea
\mathfrak{I}_{ 0}^{(1)} &=& \tfrac{1}{\Theta^{3/2}},\;\; 
\mathfrak{I}_{ 1}^{(1)} = - \tfrac{4\, |\omega|^{2}} {\Theta^{3/2}} 
\lg 1  -  \tfrac{3} {\Theta} \rg, \;\;
\mathfrak{I}_{ 2}^{(1)} =  \tfrac{4} {\Theta^{3/2}} 
\lg 1  -  \tfrac{36\, |\omega|^{4}} {\Theta}
-  \tfrac{24 \,\zeta  |\omega|^{4}} {\Theta}
+  \tfrac{60 \,|\omega|^{4}} {\Theta^{2}} \rg, \nn\\
\mathfrak{I}_{ 3}^{(1)} &=& - \tfrac{48\, |\omega|^{2}} {\Theta^{3/2}}
\lg 3  -  \tfrac{9 } {\Theta}
-  \tfrac{12\, |\omega|^{4}} {\Theta} - \tfrac{24 \,\zeta  |\omega|^{4}} {\Theta}
+ \tfrac{120 \, |\omega|^{4}} {\Theta^{2}} + \tfrac{120\, \zeta |\omega|^{4}} {\Theta^{2}}
- \tfrac{140 \, |\omega|^{4}} {\Theta^{3}}\rg.
\label{I-1-n}
\eea
The equations (\ref{exp-n}, \ref{I123}, \ref{I-1-n}) now jointly assemble the photon numbers in the said states:
\bea
{}_{0}\langle \psi(t)|\mathfrak{n}|\psi(t)\rangle_{0} &=& \tfrac{1}{\Theta} - 1, \qquad 
{}_{1}\langle \psi(t)|\mathfrak{n}|\psi(t)\rangle_{1} = \tfrac{3}
{\Theta} - 2, \nn\\
{}_{2}\langle \psi(t)|\mathfrak{n}|\psi(t)\rangle_{2} &=& \tfrac {\Theta^{2}  -  36\, |\omega|^{4}\, \Theta
- 24\, \zeta  |\omega|^{4}\, \Theta
+ 60\, |\omega|^{4} } { \Theta  \big( \Theta^{2} -  4 \,|\omega|^{4}\, \Theta
- 8\, \zeta  |\omega|^{4} \,\Theta
+ 12\, |\omega|^{4} \big)} - 1,\nn\\
{}_{3}\langle \psi(t)|\mathfrak{n}|\psi(t)\rangle_{3} &=& \tfrac {-3\, \Theta^{3} + 9 \,\Theta^{2} + 12\, |\omega|^{4}\, \Theta^{2} + 24 \,\zeta  |\omega|^{4}\, \Theta^{2}
 - 120\, |\omega|^{4} \Theta \,-  120\, \zeta  |\omega|^{4}\, \Theta + 140\, |\omega|^{4} } { \Theta  \big(3 \,\Theta^{2} -  12\, |\omega|^{4}\, \Theta
- 24\, \zeta  |\omega|^{4} \,\Theta
+ 20\, |\omega|^{4} \big)} - 1,
\label{n-expl}
\eea
where we observe that the null interaction time limits of various parameters described in the context of (\ref{st-t-0}) now set the characteristic behavior ${}_{n}\langle \psi(t)|\mathfrak{n}|\psi(t)\rangle_{n}\big|_{t \rightarrow 0} \rightarrow n$ (Fig. \ref{expt_n1_fig} (a, b)). 

\par

\begin{figure}
\begin{center}
\captionsetup[subfigure]{labelfont={sf}}
\subfloat[]{\includegraphics[scale=0.5,trim= 0 0 0 10,clip]{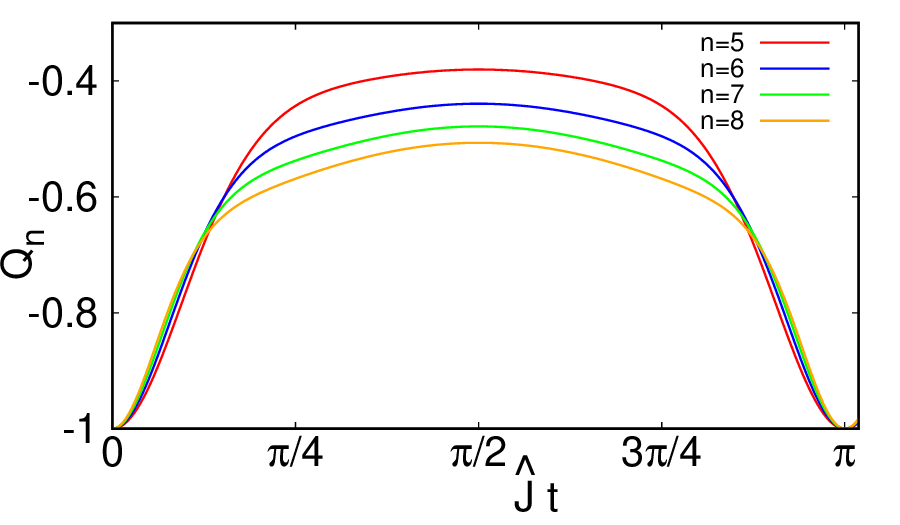}} \quad 
\subfloat[]{\includegraphics[scale=0.5,trim= 0 0 0 10,clip]{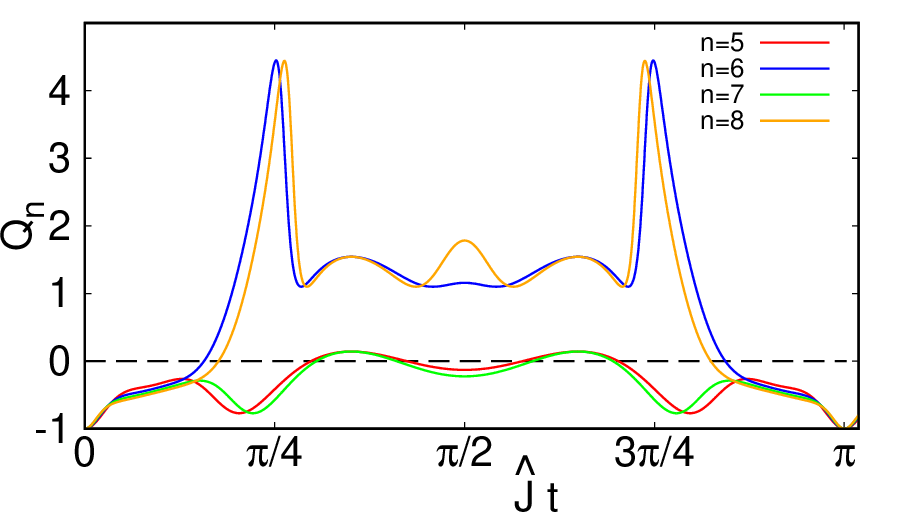}} 
\caption{ We plot the Mandel parameter $Q_{n}$ for (a) a weak  coupling $J = 0.2\, \Delta$, and (b) an ultrastrong coupling $J = 1.3\, \Delta$ cases.  For an intermediate interaction time period in diagram (a) we observe that the  excited states (\ref{n-state}) with increasing  value of $n$ display incrementally  more sub-Poissonian photon statistics. For the strong coupling case (b), however, the even parity states reveal enhanced super-Poissonian statistics, whereas their odd parity analogs possess mostly nonclassical sub-Poissonian properties.}
\label{mandel_fig}
\end{center}
\end{figure}
The second moment  outlined in (\ref{sum-I-k}, \ref{diff-I-k}) may also be provided via a similar mechanism:
\beq
\mathfrak{ I}^{(2)} (\tau) = \tfrac{1}{\sqrt{\mathfrak{D}_{+}^{\tau}\, \mathfrak{D}_{-}^{\tau}}}
\lg \tfrac{3}{4} \lo\tfrac{1+ 2 \tau}{\mathfrak{D}_{+}^{\tau}}\rc^{2} +   \tfrac{ 1- 4 \tau^{2}}{2\,\mathfrak{D}_{+}^{\tau}\, \mathfrak{D}_{-}^{\tau}} + \tfrac{3}{4} \lo\tfrac{1- 2 \tau}{\mathfrak{D}_{-}^{\tau}}\rc^{2}\rg
\equiv \sum_{n=0}^{\infty} \mathfrak{I}^{(2)}_{n} \;\tfrac{\tau^{n}}{n!}.
\label{ I2-sum}
\eeq
The power series expansion in the formal variable $\tau$ now readily furnishes the matrix elements   (\ref{I-n-int}) corresponding to the second moment in an ordered fashion. Employing the bivariate polynomials (\ref{prod-pol}) we may express an arbitrary 
expansion coefficient as follows:
\bea
\mathfrak{ I}^{(2)}_{ n} &=& \tfrac{2^{n}} {\sqrt{\Theta}} \lg \tfrac{3}{4\,(1-2 \zeta |\omega|^{2})^{2}} \ls  \mathrm{P}_{n} \lg \tfrac{5}{2}, \tfrac{1}{2};  \mathsf{x}_{1}, \mathsf{x}_{2}\rg
 +2  n \,\mathrm{P}_{n-1} \lg \tfrac{5}{2}, \tfrac{1}{2}; \mathsf{x}_{1}, \mathsf{x}_{2}\rg +\!  n (n-1)  \times\right.\right.\nn\\
&\times& \left. \mathrm{P}_{n-2}\! \lg \tfrac{5}{2}, \tfrac{1}{2}; \mathsf{x}_{1}, \mathsf{x}_{2}\rg\rs + \tfrac{1}{2\, \Theta}  \ls  \mathrm{P}_{n} \lg \tfrac{3}{2}, \tfrac{3}{2};  \mathsf{x}_{1}, \mathsf{x}_{2}\rg
 - n(n-1) \,\mathrm{P}_{n-2} \lg\tfrac{3}{2}, \tfrac{3}{2}; \mathsf{x}_{1}, \mathsf{x}_{2}\rg \rs \nn\\
&+&\!\!\!\left. \tfrac{3}{4\,(1+2 \zeta |\omega|^{2})^{2}}\! \ls  \!\mathrm{P}_{n} \!\lg\! \tfrac{1}{2}, \tfrac{5}{2};  \mathsf{x}_{1}, \mathsf{x}_{2}\rg\!
- \! 2  n \mathrm{P}_{n-1}\! \lg \!\tfrac{1}{2}, \tfrac{5}{2}; \mathsf{x}_{1}, \mathsf{x}_{2}\rg\! \!+\!  n (n-1\!)\mathrm{P}_{n-2}\! \lg \!\tfrac{1}{2}, \tfrac{5}{2}; \mathsf{x}_{1}, \mathsf{x}_{2}\!\rg\!\rs\!\rg.
\label{I2n-explicit}
\eea
Its elementary examples are noted herein:
\bea
\mathfrak{ I}^{(2)}_{ 0} &=& -  \tfrac{1}{\Theta^{3/2}} \lg 1  - \tfrac{3}{\Theta}\rg, \qquad
\mathfrak{ I}^{(2)}_{ 1} =  -  \tfrac{12\, |\omega|^{2}}{\Theta^{5/2}} \lg 3  - \tfrac{5}{\Theta}\rg, \nn\\
\mathfrak{ I}^{(2)}_{ 2} &=&  -  \tfrac{4}{\Theta^{3/2}} \lg 1 - \tfrac{3}{\Theta} - \tfrac{36\, |\omega|^{4} }{\Theta}  - \tfrac{24\,\zeta |\omega|^{4} }{\Theta}
+ \tfrac{360\, |\omega|^{4}}{\Theta^{2}} +  \tfrac{120\, \zeta |\omega|^{4} }{\Theta^{2}} - \tfrac{420\, |\omega|^{4} }{\Theta^{3}}\rg, \nn\\
\mathfrak{I}_{3}^{(2)} &=& - \tfrac{48 \,|\omega|^{2}}{\Theta^{5/2}} \lg 27 - \tfrac{45}{\Theta} - \tfrac{300 \, |\omega|^{4}}{\Theta}
 - \tfrac{360 \, \zeta |\omega|^{4}}{\Theta} +\tfrac{1400\,  |\omega|^{4}}{\Theta^{2}}
+ \tfrac{840 \, \zeta |\omega|^{4}}{\Theta^{2}} - \tfrac{1260 \, |\omega|^{4}}{\Theta^{3}} \rg.
\label{I2-n}  
\eea  
The photon number variance for the post-measurement states now follows  from the identity                  
\beq
{}_{n}\langle \psi(t)|\mathfrak{n}^{2}|\psi(t)\rangle_{n} = \tfrac{\mathfrak{I}^{(2)}_{ n}}{\mathfrak{I}^{(0)}_{n}} 
 - 3 \, \tfrac{\mathfrak{I}^{(1)}_{ n}}{\mathfrak{I}^{(0)}_{n}} + 1,
\label{exp-n2}
\eeq
and the  expectation value (\ref{exp-n}):
\beq
{}_{n}\langle \psi(t)| ( \Delta \mathfrak{n})^{2}|\psi(t)\rangle_{n} = \tfrac{\mathfrak{I}^{(2)}_{ n}}{\mathfrak{I}^{(0)}_{n}} -  \tfrac{\mathfrak{I}^{(1)}_{ n}}{\mathfrak{I}^{(0)}_{n}}
 \lg \tfrac{\mathfrak{I}^{(1)}_{ n}}{\mathfrak{I}^{(0)}_{n}} +1 \rg.
\label{n-var}
\eeq
Its explicit values for first few states may be obtained via the lists (\ref{I123}, \ref{n-expl}, \ref{I2-n}). The $t \rightarrow 0$ limit expressed in  the discussions relating to (\ref{st-t-0}) now yields  null value for the variance of the number operator thereat: ${}_{n}\langle \psi(t)| ( \Delta \mathfrak{n})^{2}|\psi(t)\rangle_{n}\big|_{t \rightarrow 0} \rightarrow 0$. We note that the above limits on the photon density and its variance are consistent with the observation (\ref{st-t-0}). The compendium of above results now allows us to extract the Mandel parameter (\ref{mandel-def}) for these states. As the corresponding expressions are voluminous we only provide their diagrammatic representations. 

\begin{figure}
\begin{center}
\captionsetup[subfigure]{labelfont={sf}}
\subfloat[]{\includegraphics[scale=0.5,trim= 0 0 0 10,clip]{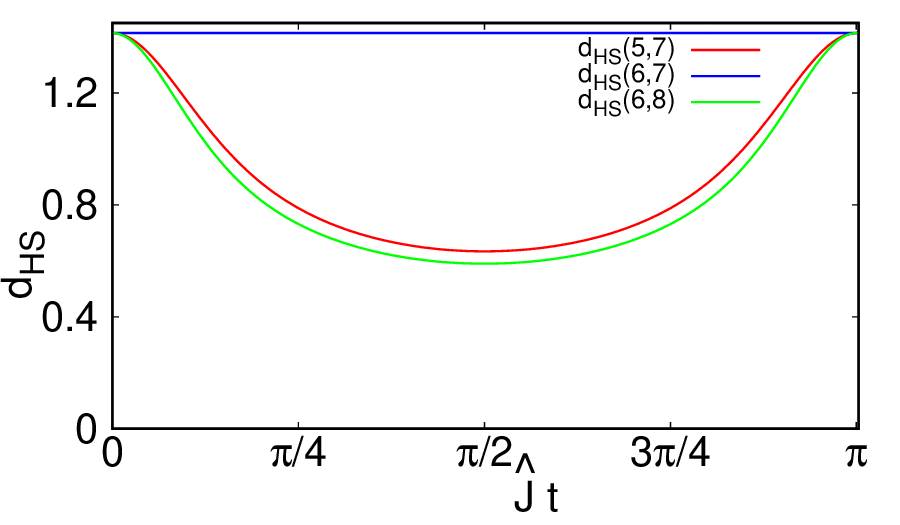}} \quad 
\subfloat[]{\includegraphics[scale=0.5,trim= 0 0 0 10,clip]{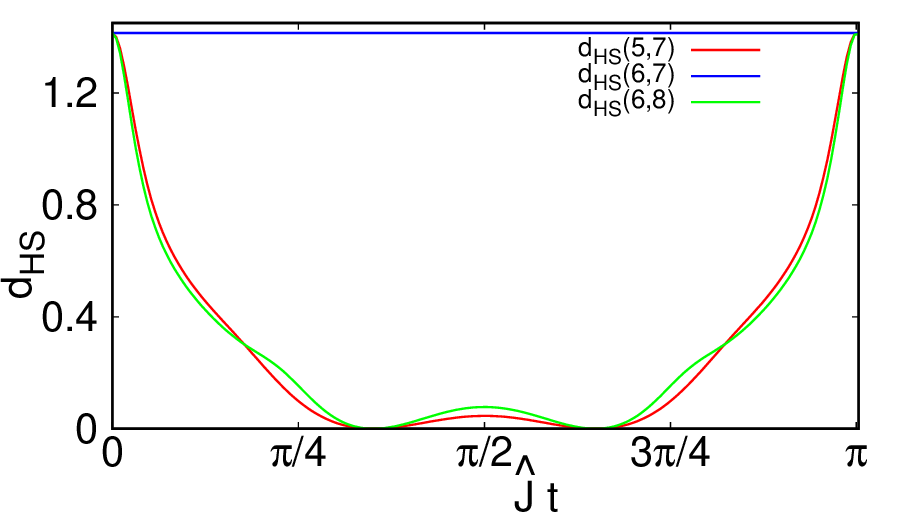}} 
\caption{The Hilbert-Schmidt distances $\mathrm{d_{HS}}$ are plotted for the cases in point: (a) $J = 0.2\, \Delta$, and (b)  $J = 1.3\, \Delta$. (a) In the weak coupling domain the distances between distinct parity states are of the same order $\mathrm{d_{HS}}(6, 8), \mathrm{d_{HS}}(5, 7) \lesssim \mathrm{d_{HS}}(6, 7)$, and the physical properties of the states in both segments are similar.  (b) For the ultrastong coupling case at the intermediate interaction time range the states separate into an even and an odd parity sectors. This is evident from the ordering in this time domain:  $\mathrm{d_{HS}}(6, 8), \mathrm{d_{HS}}(5, 7) \ll \mathrm{d_{HS}}(6, 7)$. Consequently, the nonclassicality indicators  strongly diverge for these two groups.}
\label{hs_diatance_fig}
\end{center}
\end{figure}
\par

The expectation values of the number operator and the corresponding Mandel parameter are plotted in Figs. \ref{expt_n1_fig} (a, b), and \ref{mandel_fig} (a, b), respectively. At the $t \rightarrow 0$ boundary value, and its periodic analog $\widehat{J}t \rightarrow \pi$, the diagrams represent the limiting composition (\ref{st-t-0}) of the post-measurement state, which, in turn, induces realization of
 maximal sub-Poissonian photon statistics related to the Fock states: $Q_{n}|_{t \rightarrow 0}  \rightarrow -1$.  In the weak coupling regime, say at $J = 0.2\, \Delta$ (Fig. \ref{mandel_fig} (a)), the higher values of the quantum number $n$ indicate more nonclassicality as the corresponding Mandel parameters display an increased sub-Poissonian property.  
Moreover, reciprocating the characteristic observed in the context of the Wigner $\mathrm{W}$-distribution (Sec. \ref{wigner}) the Mandel parameter $Q_{n}$ in this regime 
(Fig. \ref{mandel_fig} (a)) maintains a quasi-stationary value in the intermediate interaction time domain $\widehat{J}t \sim \tfrac{\pi}{2}$. This is in contrast to its behavior in the 
$\widehat{J}t \ll1$ regime and its cyclic counterparts. As previously noted the quasi-stationary property facilitates possible experimental use of these states as a source of a controlled level of nonclassicality.  In the  ultrastrong coupling regime, given, for instance, by the choice $J = 1.3\, \Delta$, the expectation value of the number operator, as well as the Mandel parameter, tend to \textit {coalesce together in two distinct orbits distinguished by their parity eigenvalues}. To demonstrate the emergence of distinct parity  dependent nonclassicalty in the strong coupling domain we utilize Hilbert-Schmidt metric elements (\ref{HSM}) corresponding to relevant states.  In Fig. \ref{hs_diatance_fig} (a, b) we plot two metric elements, one between the parity even states (say, $\mathrm{d_{HS}}\lo 6, 8\rc$), and another between the odd states
(say, $\mathrm{d_{HS}}\lo 5, 7\rc $). A cross metric element  linking an  even  and an odd state shows maximal Hilbert-Schmidt distance: for instance, $ \mathrm{d_{HS}}\lo 6, 7\rc = \sqrt{2}$, as, following (\ref{even-odd}), these states are mutually orthogonal. Excluding the $t \rightarrow 0$ limit and its periodic counterparts, where superposition of Fock states is not appreciable,  we, in the ultrastrong coupling example, observe (Fig. \ref{hs_diatance_fig} (b)) an ordering $\mathrm{d_{HS}}\lo 6, 8\rc, \mathrm{d_{HS}}\lo 5, 7\rc \ll \mathrm{d_{HS}}\lo 6, 7\rc$ that strongly suggests separation of the states in two classes marked by their parity eigenvalues. This feature is further emphasized by the observations made in the context of Fig. \ref{hs_diatance_even_odd} in Appendix B.
In the ultrastrong coupling realm we notice (Fig. \ref{mandel_fig} (b)) that the parameter $Q_{n}$ remains mostly  negative (positive) for the parity odd (even) states rendering the
corresponding photon statistics sub-Poissonian (super-Poissonian). Barring the initial time and its cyclic analogs, in this coupling regime  the   state (\ref{n-state})  incorporates superposition of a large number of Fock states in a particular parity segment. Therefore, the  set of states (\ref{n-state}) intrinsically  splits into two parity-specific parts, where all states within a sector have largely identical nonclassicality structures.
In addition we note that regarding the photon statistics, the ensemble of states (\ref{n-state})  has qualitative similarity with the even  and the odd Schr\"{o}dinger cat states [\cite{A2013}]. Another evident feature (Fig. \ref{expt_n1_fig} (b)) is that in the ultrastrong coupling domain the expectation value of the occupation number of photons approaches, at certain times in the oscillatory cycle,  a lower limit consistent with the parity characteristics of the state: 
${}_{n}\langle \psi(t)|\mathfrak{n}|\psi(t)\rangle_{n} \rightarrow 0 (1)$ for the even (odd) values of   $n$.   

\section{Interaction time dependent squeezing}
\label{squeeze}
\begin{figure}
\begin{center}
\captionsetup[subfigure]{labelfont={sf}}

\subfloat[]{\includegraphics[scale=0.6,trim= 0 0 52 30,clip]{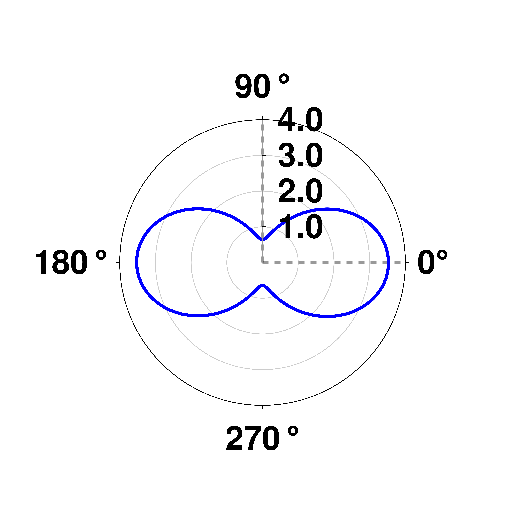}} \quad 
\subfloat[]{\includegraphics[scale=0.6,trim= 50 20 50 30,clip]{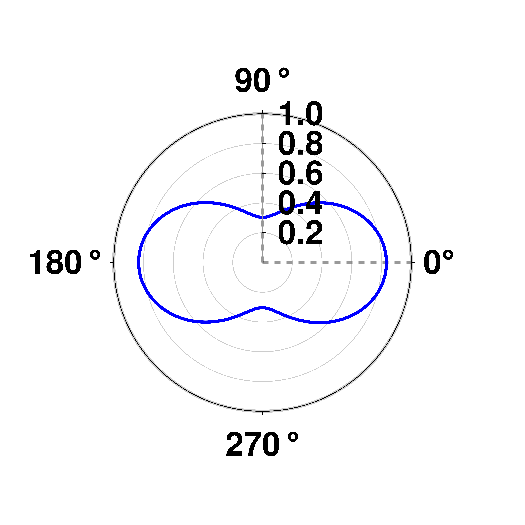}} \quad 
\subfloat[]{\includegraphics[scale=0.6,trim= 53 20 52 30,clip]{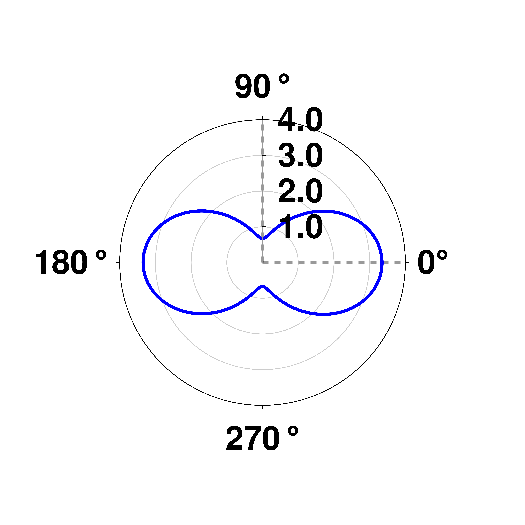}}  \quad 
\subfloat[]{\includegraphics[scale=0.6,trim= 53 20 30 30,clip] {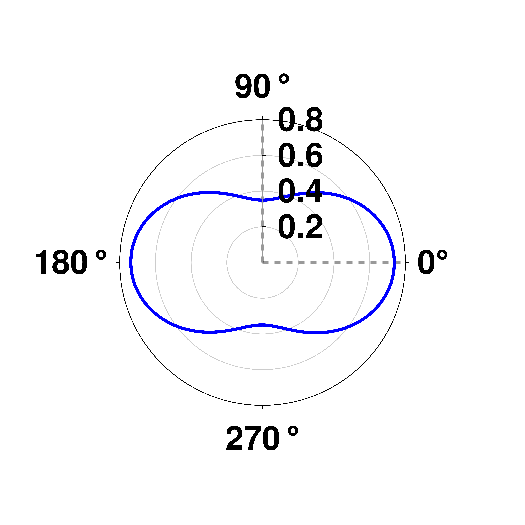}}
\caption{ The variance of the quadrature variable $\mathrm{V}_{n}(\theta)$ is plotted for an ultrastrong coupling strength $J=1.3\, \Delta$ at the interaction time $\widehat{J} t = \tfrac{\pi}{2}$ for the quantum numbers $n=5-8$, consecutively,  in diagrams (a)-(d). The maximum and minimum values of the variance occur at angles $0^{\circ}$ and ${90}^{\circ},$ respectively. The  minimum values of the variance for the said quantum numbers $n=5-8$ read, sequentially, as 0.6381, 0.3037, 0.6725, 0.3498. The corresponding  maximum values of the variance for the same ordered set of quantum numbers are  3.5265, 0.8326, 3.3460, 0.7379.}
\label{squeezed_j13}
\end{center}
\end{figure}

\setcounter{equation}{0}
The phase space $\mathrm{W}$-distributions observed for various coupling coefficients in Figs. \ref{Wigner_j02}, \ref{Wigner_j13} display signatures of squeezing that is dynamically generated in the post-measurement state (\ref{n-state}) via the interaction between the modes. It disappears  in the null interaction time limit $t \rightarrow 0$.  
To study the interaction time dependent squeezing of the post-measurement state (\ref{n-state}) we now consider the quadrature variable: $\mathrm{X}_{\theta} = \tfrac{1}{\sqrt{2}} (a \exp (-i \theta) + a^{\dagger} \exp (i \theta))$.
The coordinate and momentum operators are given, respectively, by $\mathsf {q} \equiv \tfrac{1}{\sqrt{2}}(a + a^{\dagger}) = \mathrm{X}_{0}, \mathsf {p} \equiv \tfrac{1}{i\,\sqrt{2}}(a - a^{\dagger}) = \mathrm{X}_{\tfrac{\pi}{2}}$. The parity property (\ref{even-odd}) constrains the first  moment: ${}_{n}\langle \psi(t)|\mathrm{X}_{\theta}|\psi(t)\rangle_{n} = 0$.
The corresponding second moment reads
\bea
 {}_{n}\langle \psi(t)|\mathrm{X}_{\theta}^{2}|\psi(t)\rangle_{n} &=& \tfrac{1}{2} \;{}_{n}\langle \psi(t)|\lg a^{2} \exp (- 2 i\theta) + aa^{\dagger} + a^{\dagger} a +
a^{\dagger\,2} \exp ( 2 i\theta)\rg |\psi(t)\rangle_{n}\nn\\
&=& \tfrac{1}{2 \mathfrak{I}_{n}^{(0)}} \lg \mathcal{I}_{n} \exp (- 2 i\theta) +  \mathcal{I}_{n}^{*} \exp (2 i\theta) + 2 \mathfrak{I}_{n}^{(1)}\rg - \tfrac{1}{2},
\label{squeeze-2}
\eea
where the Gaussian complex phase space integral has the following form:
\beq
\mathcal{I}_{n} = \int \tfrac{\mathrm{d}^{2} \alpha}{\pi} \, \exp \lg  - |\alpha|^{2} + \Omega^{*} \alpha^{2} 
+ \Omega \alpha^{*\,2} \rg \alpha^{2}  H_{n}\big(\omega^{*} \alpha\big)\,
 H_{n}\big(\omega  \alpha^{*}\big).
\label{a_2}
\eeq
Our  evaluation of the integral (\ref{a_2}) hinges on employing the bilocal generating function  (\ref{bilocal}), and closely parallels the description given in Secs.  \ref{state} and \ref{nonclassicality}. We quote the result for an arbitrary $n$ as follows:
\beq
\mathcal{I}_{n} = 2^{n+1}\,\tfrac{\omega^{2}}{\Theta^{3/2}} \ls \zeta\,\mathrm{P}_{n} \lg \tfrac{3}{2}, \tfrac{3}{2};  \mathsf{x}_{1}, \mathsf{x}_{2}\rg
 - n(n-1)\, (1+ \zeta)\,  \mathrm{P}_{n-2} \lg \tfrac{3}{2}, \tfrac{3}{2};  \mathsf{x}_{1}, \mathsf{x}_{2}\rg \rs.
\label{int_a2}
\eeq 
Few preliminary instances of the above  integral (\ref{int_a2}) are given below:
\bea
\mathcal{I}_{0} &=&  2 \, \omega^{2}\,\tfrac{\zeta }{\Theta^{3/2}},\; \mathcal{I}_{1} = 24\, \omega^{2}\,\tfrac{\zeta\,|\omega|^{2}}{\Theta^{5/2}}, \;
\mathcal{I}_{2} = 8\, \omega^{2}\,\tfrac{1}{\Theta^{3/2}} \lg 4 + \zeta - \tfrac{6}{\Theta} -\tfrac{ 12 \,\zeta\,|\omega|^{4}}{\Theta} +
 \tfrac{60\,\zeta\,|\omega|^{4}}{\Theta^{2}}\rg,\nn\\
\mathcal{I}_{3} &=& 16\, \omega^{2} \,\tfrac{|\omega|^{2}}{\Theta^{3/2}} \lg 180 - \tfrac{216}{\Theta} + \tfrac{54\,\zeta}{\Theta}
- \tfrac{360\,\zeta\,|\omega|^{4}}{\Theta^{2}} + \tfrac{840\,\zeta\,|\omega|^{4}}{\Theta^{3}}\rg.
\label{a2_list}
\eea
The variance of the quadrature variable $\mathrm{V}_{n}(\theta) = {}_{n}\langle \psi(t)|\mathrm{X}_{\theta}^{2}|\psi(t)\rangle_{n}$ is minimized at an angle $\theta_{\mathrm{min}}$ that obeys  the conditions
$\tfrac{\partial \mathrm{V}_{n}}{\partial \theta}\big|_{\theta_{\mathrm{min}}} = 0, \tfrac{\partial^{2} \mathrm{V}_{n}}{\partial \theta^{2}}\big|_{\theta_{\mathrm{min}}} > 0$.
Its explicit value may be noted as 
\beq
\theta_{\mathrm{min}} = \tfrac{1}{2} \lg \tan^{\mathrm{-1}}\lo \tfrac{\mathrm{Im}\; \mathcal{I}_{n}} {\mathrm{Re}\; \mathcal{I}_{n}}\rc + p\, \pi\rg, 
\; p = 0 \;\hbox {or}\, 1.
\label{sqeeze-min}
\eeq
The maximum variance of the quadrature variable  is realized in the conjugate direction $\theta_{\mathrm{max}} = \theta_{\mathrm{min}} \pm \tfrac{\pi}{2}$ in the phase space. 
 If the said  minimum variance registers a value lower than the coherent state limit: 
$\mathrm{V}_{n}\big(\theta_{\mathrm{min}}\big) < \tfrac{1}{2}$, the state  (\ref{n-state}) is known to be  squeezed along the $\mathrm{X}_{\theta_{\mathrm{min}}}$ quadrature, and, consequently,  stretched along the variable 
$\mathrm{X}_{\theta_{\mathrm{max}}}$. As the quadrature squeezing considered here is of dynamical origin, we \textit{do not} observe it in the weak 
coupling case $J \sim 0.2 \,\Delta$ whereas in the strong and ultrastrong coupling regime squeezing effects are revealed. Accordingly, we refrain from plotting the null squeezing result for the weak interaction strength.
The polar plots of the variance $\mathrm{V}_{n}(\theta)$  for an  ultrastrong coupling example ($J=1.3 \,\Delta$) at the symmetric scaled time $\widehat{J} t = \tfrac{\pi}{2}$ for various entries of the quantum number $n$ are provided in Fig. \ref{squeezed_j13}, where we mark the minimum values of the variance and the corresponding  polar angles.  The quoted minimum values of the variance $\mathrm{V}_{n}(\theta)$ suggest that the states with even quantum numbers, namely, $n = 6, 8$ exhibit squeezing which disappears for the odd  $n = 5, 7$ states. Interestingly, we have observed in Secs. \ref{wigner} and \ref{nonclassicality} that in the ultrastrong coupling regime, and  subsequent to sufficient interaction time between the modes, the odd quantum states, {\it vis-\`{a}-vis} their even counterparts, display significantly  more pronounced nonclassical features such as negativity of the  Wigner distribution and sub-Poissonian photon statistics.  On the other hand, it is evident from Fig. \ref{squeezed_j13} that in the said interaction domain  the even quantum number states are characterized by substantial coupling time dependent squeezedness, whereas the odd states are not.  To understand this contrast in nonclassicality characteristics we plot 
the distances among the even, and the odd set of states  $\lo \mathrm{d_{HS}}\lo n=0, \mathrm{even}\rc, \mathrm{d_{HS}}\lo n=1, \mathrm{odd}\rc \rc$
in Fig. \ref{hs_diatance_even_odd} in Appendix B. From the near-null values of the Hilbert-Schmidt distances plotted therein for the intermediate interaction time domain $\widehat{J} t  \sim 1$ it is evident that even (odd) set of states form closely adjacent orbits in the Hilbert space while maintaining a large separation \textit{between} the two parity-distinct sets 
(Fig. \ref{hs_diatance_fig} (b)).  As the $n=0$ state is pure Gaussian, other $n=$ even states exhibit, in this interaction time domain,  near-Gaussian characteristics, and, consequently, the corresponding negativity of the Wigner distribution remains insignificant: $\delta_{\mathrm{W}} \ll 1$ (Fig. \ref{Wigner_j13} (b)). In contrast the even states therein display appreciable squeezing property for ultrastrong coupling strength. The absence of squeezing for the $n=$ odd states in this interaction time regime may be understood as follows. 
The states $|\psi(t)\rangle_{n=\mathrm{odd}}$, in general,  incorporate a superposition of a large number of odd parity Fock states.  In the domain of interest the Hilbert-Schmidt distance $\mathrm{d_{HS}}(1, \mathrm{odd}) \rightarrow 0$ (Fig. \ref{hs_diatance_even_odd}). The $n=$ odd states, therefore, closely  follow the characteristics of $n=1$ state at the concerned interaction time span. However, despite the said superposition effects, the variance $\mathrm{V}_{n=1}(\theta)$ never decreases below the threshold coherent state value $(\tfrac{1}{2})$, 
which, in turn, leads to a general absence of squeezing for $n=\mathrm{odd}$ states. 
\section{Conclusion}
\label{conclude}
\setcounter{equation}{0}
Considering  a system of two coupled waveguides equipped with  interacting Bosonic modes, we introduce a bimodal squeezed vacuum as an input state. After the degrees of freedom undergo a hopping interaction for a designated time $t$, a projective measurement is executed on one of the modes situating it at a Fock state. As a consequence of the nonunitary process of  measurement, certain nonclassical characteristics develop in the state of the residual degree of freedom. A post-measurement squeezed Hermite polynomial state appears in the complementary mode. The parameters of the state demonstrate periodic  dependence on the interaction time prior to the measurement.  For this general class of states the Wigner  distribution is obtained. The negativity of the Wigner distribution of the 
non-null modes points towards their nonclassicality. In the strong and ultrastrong coupling domains the ensemble of states splits into two sectors depending on their parity eigenvalues. 
The odd parity sector display prominent negativity of the Wigner distribution. We also study the non-Poissonian photon statistics of this class of states via the Mandel parameter. In the weak coupling regime sub-Poissonian photon  statistics is evidenced. Furthermore, while the parity odd states in the ultrastrong coupling realm at intermediate interaction time zone largely display negative values of the Mandel parameter signifying 
nonclassical sub-Poissonian statistics, the parity even states therein, on the other hand, exhibit predominantly positive Mandel parameter implying super-Poissonian photon statistics.  Contrarily, in the same  domain the even states are characterized by another aspect of nonclassicality, namely  the dynamically generated squeezedness that is conditioned by the time of interaction between the modes and the coupling strength. Insofar as nonclassical features are concerned, the even and odd quantum states possess complementary characteristics. These properties of the post-measurement state owe their origin in the superposition of the Fock states that maintains identical parity eigenvalues. One advantage of the nonclassical states considered here is that their extent of nonclassicality can be controlled via selecting the  time-span and strength of the interaction between the modes. Moreover, for a considerable span of interaction time the characteristics of the state remain quasi-stationary. These features may be useful in experimental application of these states.

\par

As a continuation of the present work  we wish to consider an array of coupled waveguides furnished with a  multi-mode initial state. Interestingly, using waveguide elements subject to a parabolic coupling distribution, coherent transfer of quantum states has been experimentally achieved [\cite{Perez2013}]. Integrated devices such as waveguide arrays with electro-optically controllable Hamiltonian parameters has recently been established [\cite{Yang2024}]. In such systems endowed with multiple interacting Bosonic degrees of freedom, sequential measurements on individual modes are likely to generate  fruitful multi-entangled states with cumulative nonclassicality.

\section*{Acknowledgments}
One of us (RC) acknowledges the hospitality in the Department of Nuclear Physics, and  the Central Instrumentation Research and Service Laboratory,  University of Madras, where part of the work has been done. This work has been partially supported by the grant DST-SERB  ref. CRG/2020/004323, Government of India.

\section*{Appendix A}
\label{appendix-Wigner}
\renewcommand{\theequation}{A-\arabic{equation}}
\setcounter{equation}{0} 

Following (\ref{W-n-gen}) here we quote the first few examples of the Wigner distribution for the $n$-th 
post-measurement state  (\ref{n-state}) described by the phase space variables:
\bea
\mathrm{W}_{0}(\alpha, \alpha^{*}) &=&\!\! \tfrac{1}{\mathfrak{I}_{0}^{(0)}}\; \mathcal{G}(\alpha, \alpha^{*}), \;
\mathrm{W}_{1}(\alpha, \alpha^{*}) = - \tfrac{4\,|\omega|^{2}}{\mathfrak{I}_{1}^{(0)}}  \mathcal{G}(\alpha, \alpha^{*}) \!\!
\lg \tfrac {1+ 4 |\alpha|^{2}}{\Theta } - 8\, \tfrac{|\alpha|^{2} -\zeta\;\langle \omega,\alpha \rangle_{2}}{\Theta^{2}}\rg, 
\nn\\
\mathrm{W}_{2}(\alpha, \alpha^{*}) &=& \!\!\!\! \tfrac{1}{\mathfrak{I}_{2}^{(0)}} \; \mathcal{G}(\alpha, \alpha^{*}) 
\lg 4 - \tfrac {16}{\Theta } \ls \big(1+ 2 \zeta\big) |\omega|^{4} 
- 2 \, \langle \omega,\alpha \rangle_{2} \rs + \tfrac {16}{ \Theta^{2}} \;\; \times\right.\nn\\
&& \!\!\!\!\times  \ls  \lo 3+ 32 |\alpha|^{2} + 16 \zeta |\alpha|^{2} - 8 \zeta  \langle \omega,\alpha \rangle_{2} 
+  16 |\alpha|^{4} \rc |\omega|^{4} - 4 \, \langle \omega,\alpha \rangle_{2} \rs \nn\\
&& \!\!\!\! - \tfrac {256}{\Theta^{3}}  \ls  \lo  3  |\alpha|^{2} + 6  |\alpha|^{4}\! - \! 3 \zeta \langle \omega,\alpha \rangle_{2} \! - 4 \zeta  |\alpha|^{2}
\langle \omega,\alpha \rangle_{2}  \rc \! |\omega|^{4}\!+\!\langle \omega,\alpha \rangle_{4} \rs  \nn\\
&&\left. \!\!\!\! + \tfrac {256}{\Theta^{4}}  \ls  2\lo  3  |\alpha|^{4} 
- 4 \zeta |\alpha|^{2} \langle \omega,\alpha \rangle_{2}  \rc |\omega|^{4} + \langle \omega,\alpha \rangle_{4} \rs \rg,\nn\\
\mathrm{W}_{3}(\alpha, \alpha^{*}) &=& \!\!\!\!- \tfrac{48\,|\omega|^{2}}{\mathfrak{I}_{3}^{(0)}} \; \mathcal{G}(\alpha, \alpha^{*})  
\sum_{\jmath = 1}^{6} \tfrac{\mathcal{T}_{\jmath}}{\Theta^{\jmath}}.
\label{W123}
\eea
The cited coefficients in the last equality in (\ref{W123})  read as follows:
\bea
\mathcal{T}_{1} \!\!\! &=& \!\!\! 3 \lg 1+ 4 |\alpha|^{2}\rg,\nn\\ 
\mathcal{T}_{2}\!\!\! &=&\!\!\! -12 \lg 1+2 \zeta + 4 (1 + 2 \zeta) |\alpha|^{2}\rg  |\omega|^{4} 
- 8 \lg 3 |\alpha|^{2}- \big(9 + 3 \zeta  + 4  |\alpha|^{2}\big) \langle \omega,\alpha \rangle_{2}\rg,\nn\\
\mathcal{T}_{3} \!\!\!&=&\!\!\!  4 \lg 5 +12 (9+ 8 \zeta) |\alpha|^{2} -16 \big(3+2  |\alpha|^{2}\big) \zeta  \langle \omega,\alpha \rangle_{2}  + 48 (3+ 2 \zeta) |\alpha|^{4} 
+ \tfrac{64}{3} |\alpha|^{6} \rg |\omega|^{4}\nn\\
&&\!\!\! -32 \lg 3+8  |\alpha|^{2}\rg  \langle \omega,\alpha \rangle_{2}  +64 \lg 1+ \zeta \rg \langle \omega,\alpha \rangle_{4}, \nn\\
\mathcal{T}_{4} \!\!\!&=&\!\!\! -32 \lg 15 |\alpha|^{2} + 24  (3+  \zeta) |\alpha|^{4} + 32 |\alpha|^{6} - \big(15 + 56 |\alpha|^{2} + 16 |\alpha|^{4}\big)  
 \zeta  \langle \omega,\alpha \rangle_{2} \rg |\omega|^{4}\nn\\
&&\!\!\!+ 256  |\alpha|^{2} \langle \omega,\alpha \rangle_{2} - 128  \lg 3 + \zeta + 2  |\alpha|^{2}\rg \langle \omega,\alpha \rangle_{4}, \nn\\
\mathcal{T}_{5}\!\!\! &=&\!\!\! 640 \lg 3  |\alpha|^{4} - 4 \zeta  |\alpha|^{2} \langle \omega,\alpha \rangle_{2} + 4 |\alpha|^{6} 
 - 4  \zeta  |\alpha|^{4} \langle \omega,\alpha \rangle_{2}\rg |\omega|^{4}\nn\\
&&\!\!\!+ 64 \lg 5 + 12 |\alpha|^{2} \rg  \langle \omega,\alpha \rangle_{4} - \tfrac{512}{3} \zeta \langle \omega,\alpha \rangle_{6},\nn\\
\mathcal{T}_{6}\!\!\! &=&\!\!\!  -512 \lg \tfrac{10}{3} |\alpha|^{6} - 5 \zeta  |\alpha|^{4} \langle \omega,\alpha \rangle_{2} \rg |\omega|^{4}
- 512 \lg |\alpha|^{2} \langle \omega,\alpha \rangle_{4} - \tfrac{1}{3} \zeta  \langle \omega,\alpha \rangle_{6}\rg.
\label{3-coeff}
\eea
\renewcommand{\thefigure}{A-\arabic{figure}}

\setcounter{figure}{0}
\begin{figure}
\begin{center}
\captionsetup[subfigure]{labelfont={sf}}
\subfloat[]{\includegraphics[scale=0.36,trim= 0 0 0 38,clip]{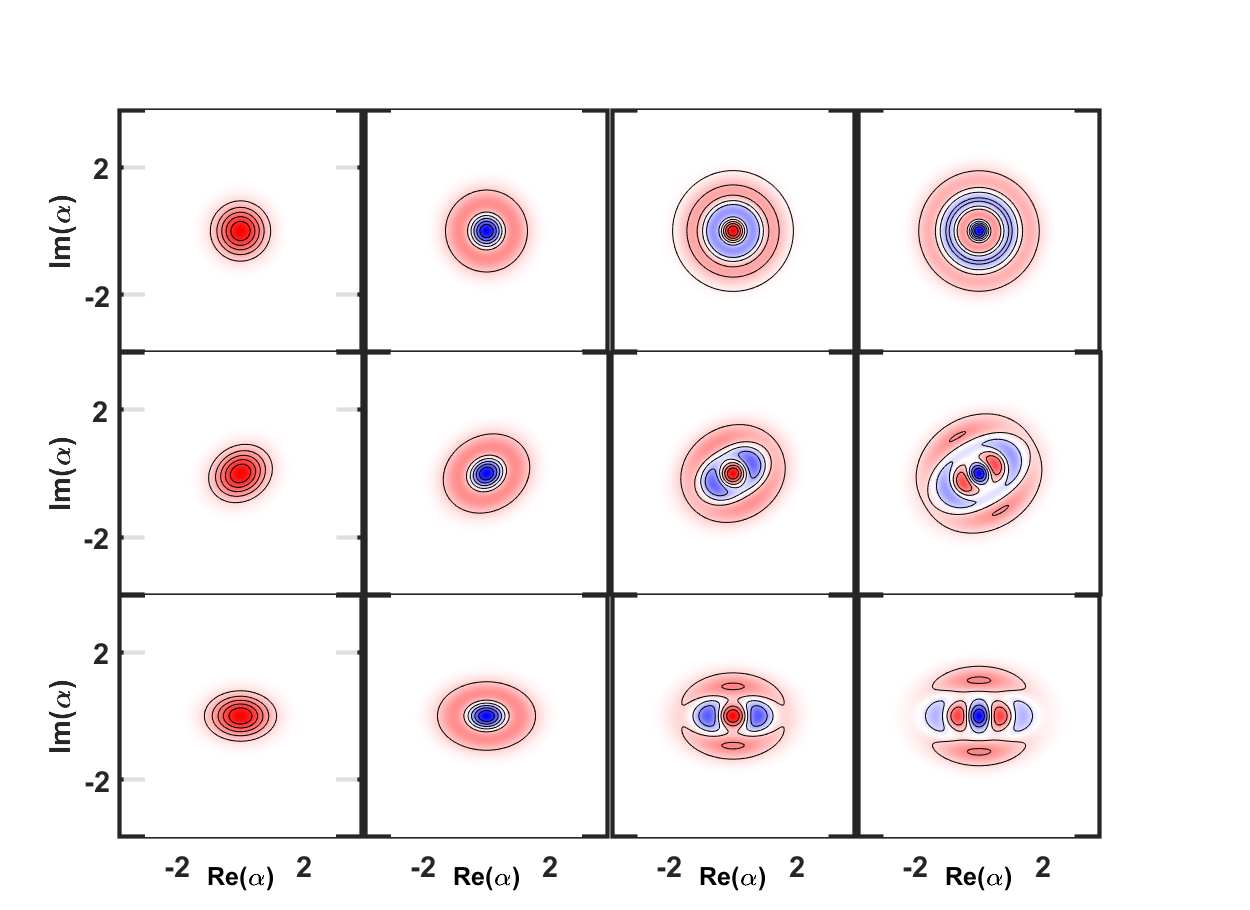}} \quad 
\subfloat[]{\includegraphics[scale=0.36,trim= 0 0 0 38,clip]{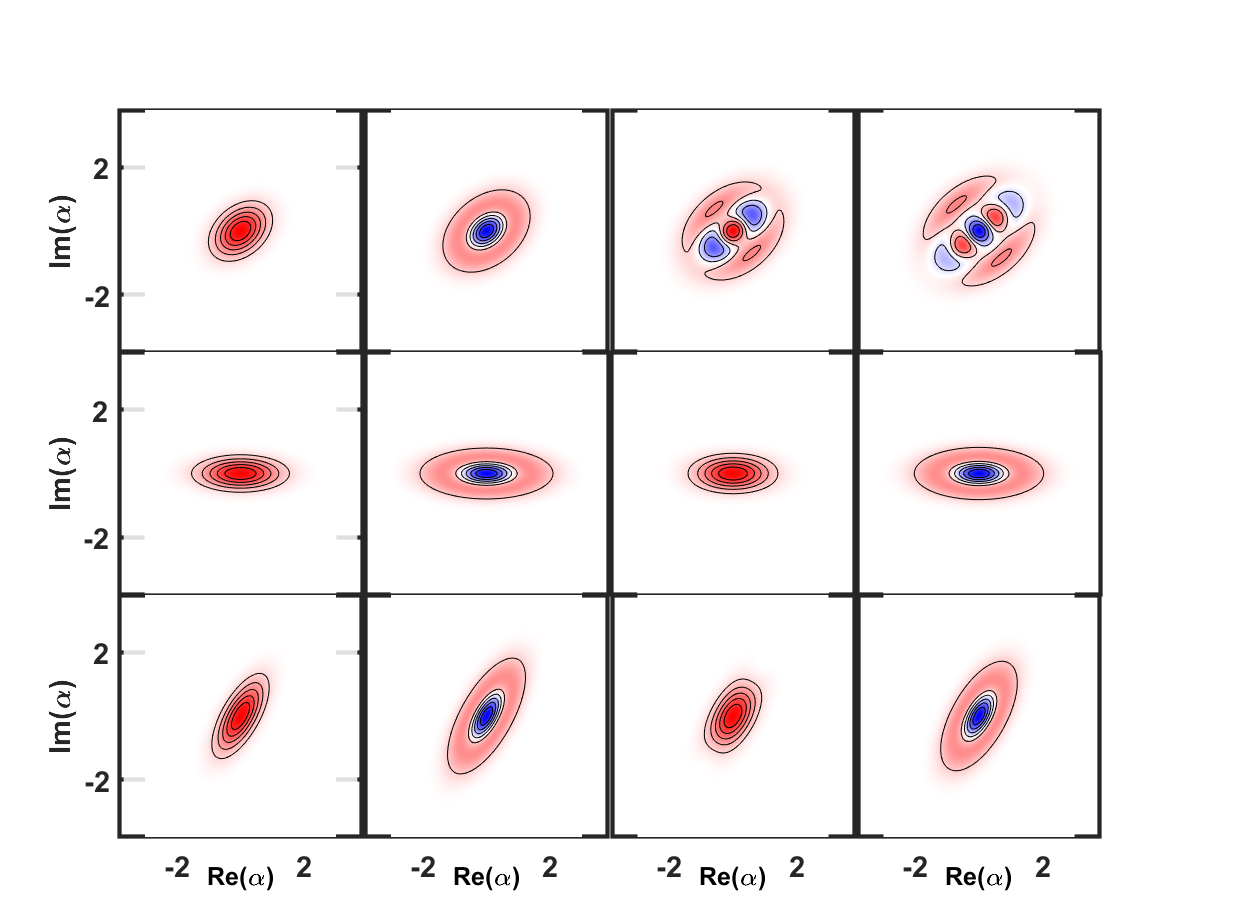}} 
\caption{(a) In the present diagram, the Wigner $\mathrm{W}$-distribution in the  weak coupling domain such as  $J = 0.2\,  \Delta$ at scaled interaction times $\widehat{J} t = 0, \tfrac{\pi}{6}, \tfrac{\pi}{2}$, and (b) ultrastrong coupling domain (say, $J = 1.3\,  \Delta$) for the choices $\widehat{J} t = \tfrac{\pi}{12}, \tfrac{\pi}{2}, \tfrac{3\pi}{4}$. The consecutive rows in the diagram depict the noted times in increasing order from top to bottom. The columns from left to right reproduce the distributions for the quantum numbers 
$n= 0-3$ in the ascending sequence.}
\label{Wigner_0123}
\end{center}
\end{figure}
The $\mathrm{W}$-distributions listed above explicitly satisfy the normalization constraint (\ref{Wigner_def_S}). As in the previous case (Sec. \ref{wigner}) we now illustrate them 
both for the weak coupling ($J = 0.2\, \Delta$) and the ultrastrong coupling ($J = 1.3\, \Delta$) instances. Their generic properties agree with our description given in the 
context of Figs. \ref{Wigner_j02} and \ref{Wigner_j13}. We note that the Wigner distribution for the $n= 0$ ground state $\mathrm{W}_{0}(\alpha, \alpha^{*})$, however, is purely Gaussian, and therefore remains a non-negative quantity on the  phase space for all interaction periods.

\section*{Appendix B}
\label{appendix-HS}
\renewcommand{\theequation}{B-\arabic{equation}}
\setcounter{equation}{0}
\renewcommand{\thefigure}{B-\arabic{figure}}
\setcounter{figure}{0}
In our previous discussions (Secs. \ref{wigner}-\ref{squeeze}) we observed that in the ultrastrong coupling regime (say, $J = 1.3\, \Delta$) the states (\ref{n-state}) with even and odd parity eigenvalues exhibit different characteristic properties. In Fig. \ref{hs_diatance_fig} (b) we noticed that in the intermediate interaction time zone $\widehat{J} t \sim 1$ the 
intra-even and intra-odd Hilbert-Schmidt distances become vanishingly small: $\mathrm{d_{HS}\lo even, even\rc}, \mathrm{d_{HS}\lo odd, odd\rc} \ll 1$. To investigate this issue further here we visualize the distances between $|\psi(t)\rangle_{n=0}$  and other $|\psi(t)\rangle_{n=\mathrm{even}}\; \big(|\psi(t)\rangle_{n=1}$  and other $|\psi(t)\rangle_{n=\mathrm{odd}}\big)$ states. In particular, we observe that the distances $\mathrm{d_{HS}}\lo n=0, \mathrm{even}\rc \rightarrow 0 \;
\big(\mathrm{d_{HS}}\lo n=1, \mathrm{odd}\rc \rightarrow 0\big)$ at certain instances at the interaction time range $\widehat{J} t \sim 1$. In the vicinity of these occurrences all even (odd) states show near-identical properties which closely resemble those of $n = 0 \,(1)$ state. For example, this feature effects near-Gaussian behavior of all even states in the concerned time periods as well as a general absence of squeezing for the $n$-odd states therein.

\begin{figure}[H]
\begin{center}
\captionsetup[subfigure]{labelfont={sf}}
\subfloat[]{\includegraphics[scale=0.5,trim= 0 0 0 10,clip]{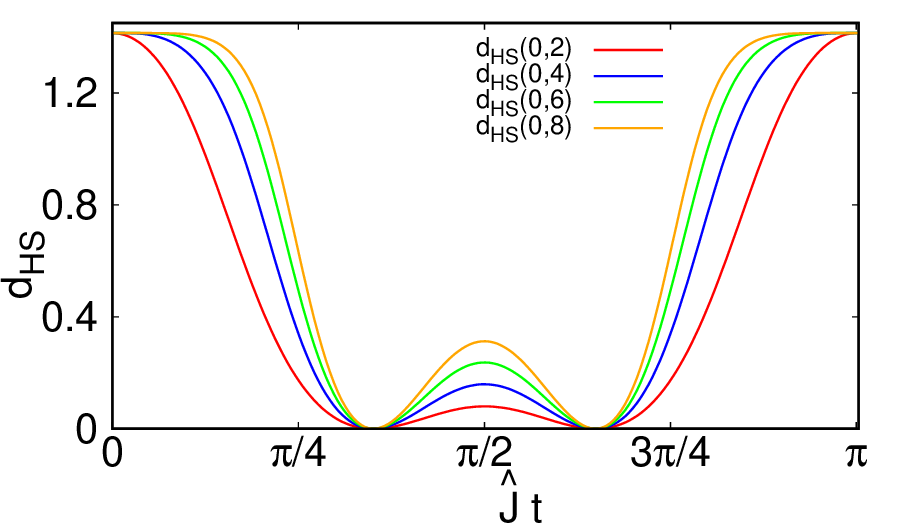}} \quad 
\subfloat[]{\includegraphics[scale=0.5,trim= 0 0 0 10,clip]{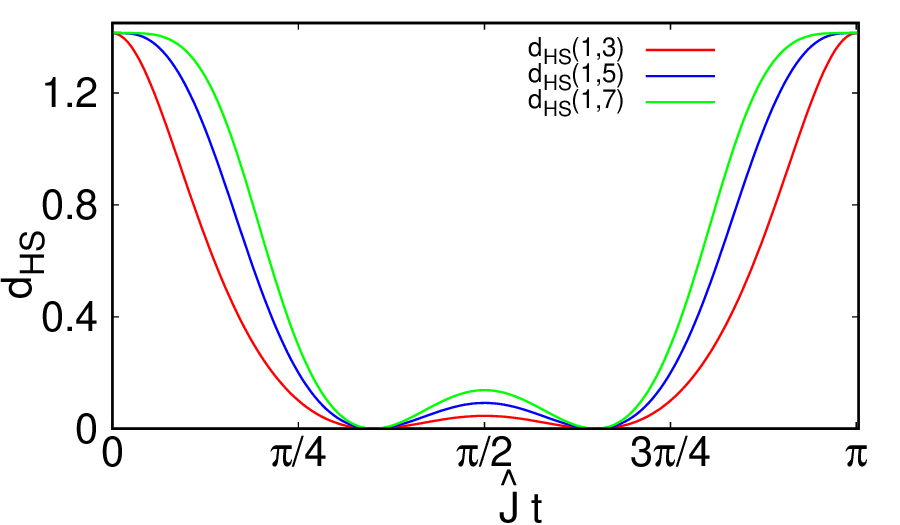}} 
\caption{The Hilbert-Schmidt distances $\mathrm{d_{HS}}$ are plotted for the ultrastrong coupling coefficient  $J = 1.3\, \Delta$ for the examples  
(a)  $\mathrm{d_{HS}}(0, \mathrm{even})$, and (b)   $\mathrm{d_{HS}}(1, \mathrm{odd})$.}
\label{hs_diatance_even_odd}
\end{center}
\end{figure}

\end{document}